\documentclass[aip]{revtex4-1}
\usepackage{subfigure}
\usepackage{bm}
\usepackage{color}
\usepackage{multirow}
\usepackage{booktabs}
\usepackage{natbib}
\usepackage[utf8]{inputenc}
\usepackage{amsmath}
\usepackage{amsfonts}
\usepackage[T1]{fontenc}
\usepackage{mathptmx}
\usepackage{algorithm}
\usepackage{algorithmic}
\usepackage{graphicx}       
\usepackage{CJKutf8}

\begin{document}
\draft 

\title{Physics-Assisted Reduced-Order Modeling for Identifying Dominant Features of Transonic Buffet}

\author{Jing Wang}
\affiliation{School of Aeronautics and Astronautics, Shanghai Jiao Tong University, Shanghai, 200240, China}	
\author{Hairun Xie}
\affiliation{Shanghai Aircraft Design and Research Institute, Shanghai, 200436, China}
\author{Miao Zhang}
\affiliation{Shanghai Aircraft Design and Research Institute, Shanghai, 200436, China}
\author{Hui Xu}
\email{dr.hxu@sjtu.edu.cn}
\affiliation{School of Aeronautics and Astronautics, Shanghai Jiao Tong University, Shanghai, 200240, China}

\begin{abstract}
Transonic buffet is a flow instability phenomenon that arises from the interaction between the shock wave and the separated boundary layer.
This flow phenomenon is considered to be highly detrimental during flight and poses a significant risk to the structural strength and fatigue life of aircraft.
Up to now, there has been a lack of an accurate, efficient, and intuitive metric to predict buffet and impose a feasible constraint on aerodynamic design.
In this paper, a Physics-Assisted Variational Autoencoder (PAVAE) is proposed to identify dominant features of transonic buffet, which combines unsupervised reduced-order modeling with additional physical information embedded via a buffet classifier.
Specifically, four models with various weights adjusting the contribution of the classifier are trained, so as to investigate the impact of buffet information on the latent space.
Statistical results reveal that buffet state can be determined exactly with just one latent space when a proper weight of classifier is chosen.
The dominant latent space further reveals a strong relevance with the key flow features located in the boundary layers downstream of shock.
Based on this identification, the displacement thickness at 80\% chordwise location is proposed as a metric for buffet prediction.
This metric achieves an accuracy of 98.5\% in buffet state classification, which is more reliable than the existing separation metric used in design.
The proposed method integrates the benefits of feature extraction, flow reconstruction, and buffet prediction into a unified framework, demonstrating its potential in low-dimensional representations of high-dimensional flow data and interpreting the "black box" neural network.
\end{abstract}



\maketitle

\section{Introduction}
Under certain ranges of angle of attack (AoA) and Mach number, shock wave oscillates vigorously and periodically over wings in flight.  
This instability phenomenon, known as transonic buffet, is caused by the interaction between the separated boundary layer and the shock wave.  
Transonic buffet develops gradually with an increasing AoA or Mach number, and the condition at which buffet first occurs is defined as buffet onset. 
The resulting shock motion and flow oscillations can have detrimental effects on aerodynamic performance, making it an undesirable phenomenon.
Certification specifications stipulate that a 1.3g margin  from the cruise operating condition to the buffet onset boundary should be maintained so as to leave a safety margin for aircraft to maneuver~\cite{timme2020global}.
Therefore, efficient and accurate prediction of the buffet onset is crucial in achieving optimal aerodynamic design~\cite{giannelis2017review}.

To address the prediction of buffet onset, researchers have explored various approaches based on computational fluid dynamics (CFD) simulations. 
Unsteady simulations, such as
unsteady Reynolds-averaged Navier-Stokes~\cite{thiery2005urans}, detached-eddy simulations~\cite{deck2005numerical}, and large-eddy simulations~\cite{dandois2018large}, have been employed to directly analyze the buffet onset.
The obtained results are in reasonable agreement with the experiments for post-critical conditions.
\citet{crouch2009origin} linked the origins of the transonic buffet to a global instability of the underlying flow, which provides good descriptors for buffet onset.
The predicted buffet onset and frequency agree well with experimental and numerical results~\cite{sartor2016mach}.
However, a wide range of AoA is required for unsteady simulations to achieve buffet conditions. Even for a moderate design space in aerodynamic design, it is computationally intensive as hundreds of simulations are required.
Industrially, a number of heuristic techniques have been developed over the years based on the global aerodynamic coefficients, such as the pitching moment break and the lift curve break method~\cite{van2012buffet}. 
A buffet-onset constraint function could be developed using global aerodynamic coefficients obtained from Reynolds Averaged Navier-Stokes (RANS) simulations, which is computationally cheaper than unsteady simulations. 
Nonetheless, implementing these methods still entails considerable computational costs since a set of flow solutions under many AoAs are required to obtain the slope of the lift curve.

Additionally, a number of empirical criteria based on the physical mechanism have also been proposed to predict the buffet onset.
One of the earliest methods is described by \citet{pearcey1958method} and \citet{pearcey1962simple},
who considered that the buffet occurs when the separation bubble extends to the trailing edge and bursts. 
\citet{redeker1976calculation} proposed a criterion based on the rear separation at 90\% of the airfoil chord. 
\citet{kenway2017buffet} proposed a separation metric based on the flow separation area, and found that a 4\% cutoff of the separation metric yields the best agreement with the lift curve break method.
Apparently, these criteria are suitable for aerodynamic optimization since they can yield fast and cheap predictions for the buffet onset.
Despite some of these criteria provide useful descriptors for the flow behaviour, they have not been effective as predictors for the buffet onset~\cite{nitzsche2009numerical,crouch2007predicting,iovnovich2012reynolds}.
The effectiveness and accuracy of empirical criteria is still a concern, which is governed by the physical mechanism that yet not fully understood~\cite{lee2001self}. 
To address the need for fast and accurate buffet analysis, \citet{li2022physics} proposed a Convolutional Neural Network (CNN) model to predict the buffet factor, which is more reliable and generalizable for different airfoil shapes. 
Although physical information, including pressure and friction distributions, is considered and taken as input instead of airfoil geometries, 
the CNN model is still hard to be interpreted and thus the reliability and generalization of the model are difficult to be proved.

The target of the present work is to identify dominant features of buffet flow, based on which a physical metric can be developed to predict the transonic buffet.
The analysis of buffet flow is challenging due to the high-dimensionality of the complex flow and the nonlinearity of the relationships within it, making the intuitive awareness of the relationships in flows very complicated~\cite{kou2018reduced}. 
Data-driven reduced-order modeling (ROM) has emerged as a powerful approach for efficiently representing and analyzing complex fluid flow phenomena~\cite{brunton2020machine, taira2017modal}. 
Methods like Proper Orthogonal Decomposition (POD)~\cite{sirovich1987turbulence}, Dynamic Mode Decomposition (DMD)~\cite{schmid2010dynamic}, and Auto-Encoders~\cite{zhang2022data, fukami2020convolutional} offer a promising alternative to directly extract essential features, reduce dimensionality and capture the underlying dynamics of the flow. 
POD captures dominant flow modes and enables efficient representation, but is limited to linear systems. 
DMD captures transient dynamics and nonlinear interactions, but it is primarily suited for unsteady analysis and has finite precision when dealing with systems that involve shock waves or abrupt changes. 
Auto-Encoders learn compact representations and handle complex data, but usually lack the interpretability and physical insights.

To address these challenges, a supervised ROM is employed to compress the high-dimensional data into a low-dimensional latent space and identify complex patterns directly without the need for manual feature extraction. 
To achieve a latent space that is highly associated with transonic buffet, a novel Physics-Assisted Variational Autoencoder (PAVAE) architecture is introduced, which integrates unsupervised reduced-order modeling with physical information via a buffet classifier. 
High-quality training samples are generated using RANS simulation with diverse geometries and flow conditions to train the model in a sample-efficient way. 
The results show that the classifier can effectively distinguish pre-buffet and post-buffet flow fields, indicating that the latent space can fully characterize transonic buffet. 
Statistical analysis is conducted to determine the contribution of each latent space to transonic buffet, and the dominant regions associated with buffet flow are identified by examining the connections between the dominant latent space and physical regions. Finally, a sensor based on the dominant region is proposed as a predictor for buffet onset.


The rest of this paper is organized as follows.
Section 2 introduces the detailed dataset involved in our study.
A brief description of the methods and detailed network used in our study are introduced in Section 3.
In Section 4, the training results and related discussions are presented. 
Finally, concluding remarks are summarized in Section 5.

\section{Database}

\subsection{Buffet onset prediction}

With the increase of AoA, transonic flow usually goes through three typical phases: weak shock waves at a low AoA (pre-buffet), strong shock waves inducing flow separations at a moderate AoA (buffet onset), and limit cycle oscillations at a large AoA (post-buffet).
For the pre-buffet phase, steady approaching method RANS can conduct credible simulations. 
For the buffet onset and post-buffet phases, unsteady and instability phenomenon are involved and RANS cannot reveal the physical dynamics of unsteady oscillations.
However, steady results can characterize the major flow structures and aerodynamic performances.
Several studies have reported that the predictions of steady RANS are consistent with flight data through the buffet-onset regime and up to near the maximum lift coefficient~\cite{kenway2017buffet,rumsey2001cfd}. 
It is clear that steady simulation is much more efficient than unsteady simulation, making it easier to simulate various geometries and flow conditions.
Therefore, RANS-based method is implemented in this study for dataset generation.

One effective alternative for transonic aircraft prediction is the use of the lift curve method among the available methods.
Under low AoA, there is no separation bubble on the upper surface and the lift increases almost linearly. 
As AoA increases, the shock-induced separation bubble decreases the lift, and a distinct slope change of the lift curve can be used as an indicator of buffet onset.
In this paper, the buffet onset point AoA$_{\text{buffet}}$ is defined at the AoA whose slope is reduced by 0.1 based on the linear part of the lift curve.
The nonlinear slope of the lift curve is calculated via interpolating the curve into a cubic spline curve. 
To reduce the impact of ambiguity, we evaluate the linear slope of the lift curve through robust regression with various combinations of AoAs, and choose the one with the highest 95\% confidence interval.

\subsection{RANS simulation}
\begin{figure}[htpb]
	\centering
	\begin{minipage}[t]{0.6\textwidth}
		\centering
		\includegraphics[width=0.49\textwidth]        {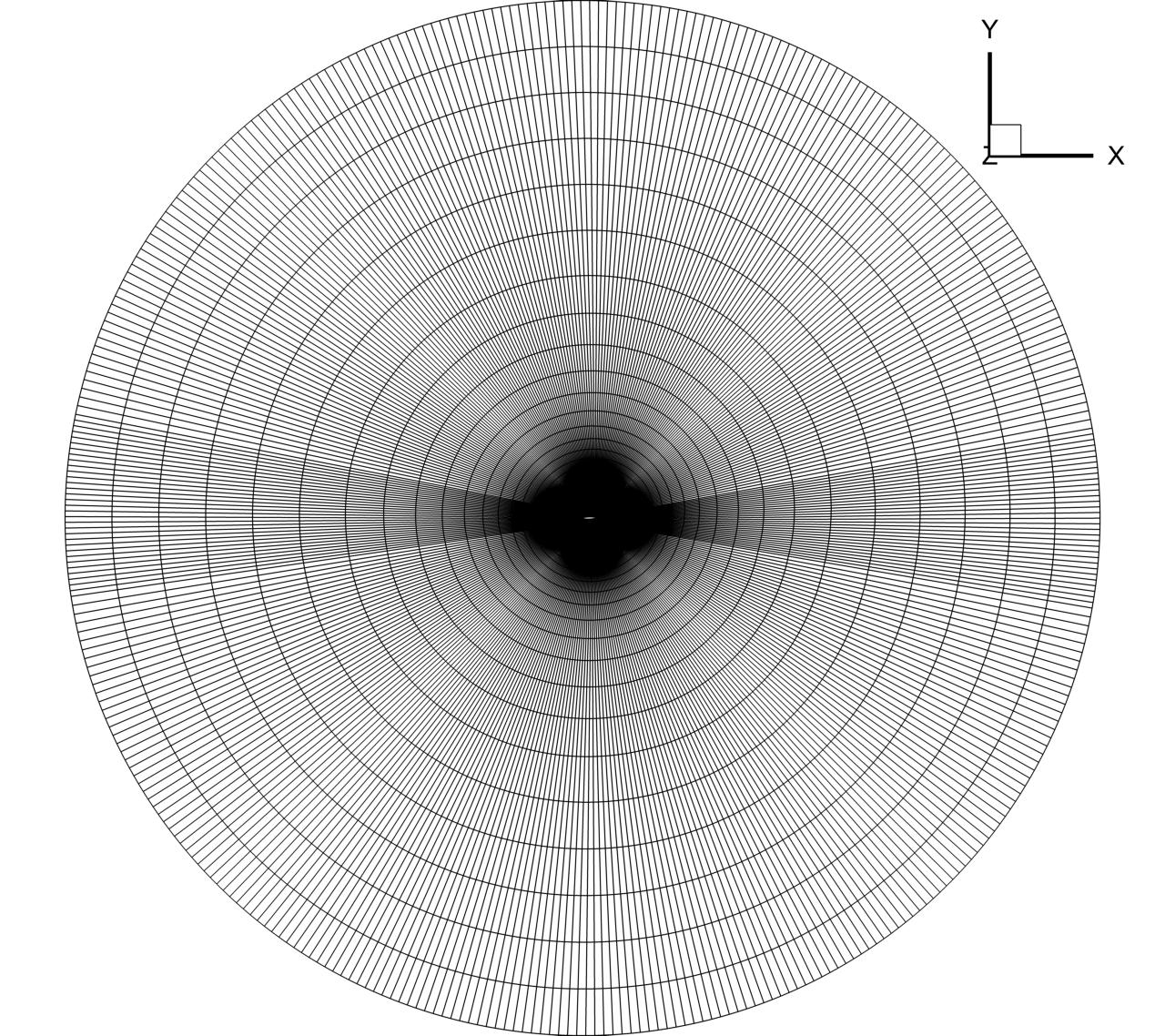}
            \includegraphics[width=0.49\textwidth]{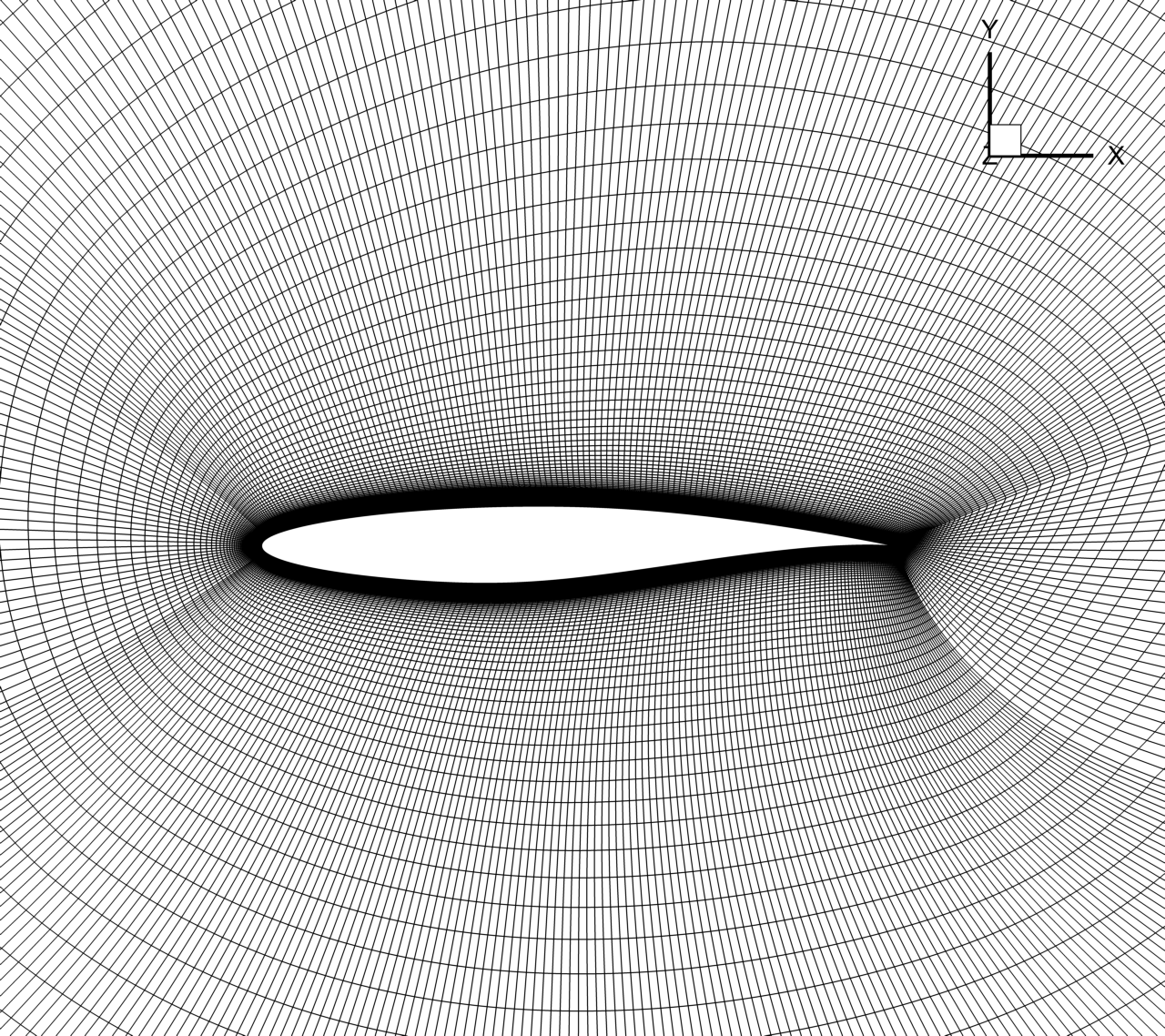}
            \caption{Outline and zoomed-in view of the structured O-mesh adjacent to the airfoil surface}
            \label{fig:mesh}
        \end{minipage}
	\hspace{10pt}
	\begin{minipage}[t]{0.35\textwidth}
		\centering
		\includegraphics[width=\textwidth]{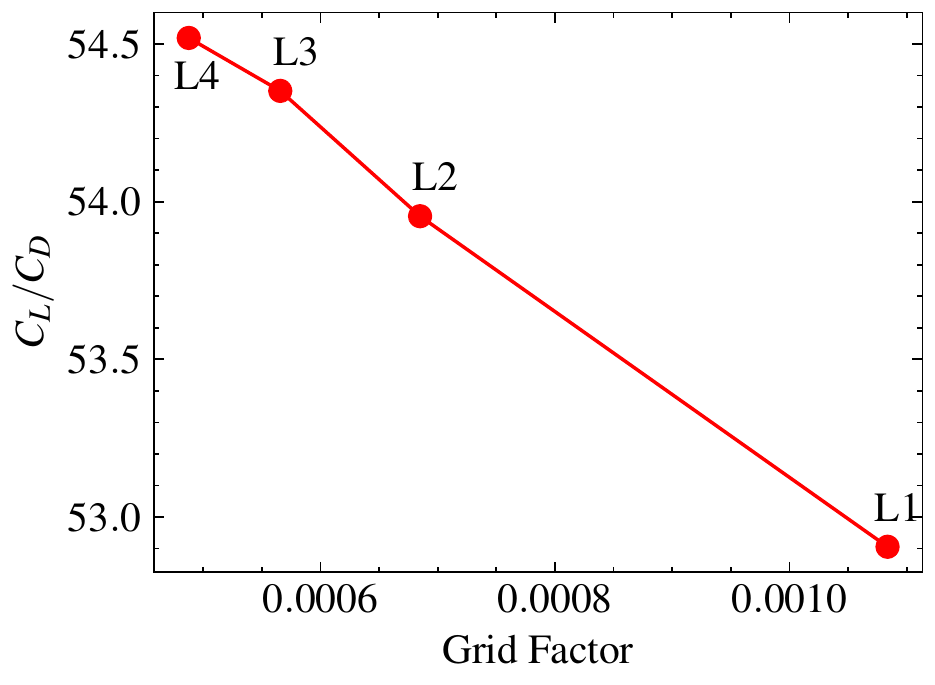}
            \caption{Results of the grid convergence study}
            \label{fig:gc}
	\end{minipage}
\end{figure}

In order to facilitate the direct representation of flow data as image pixels,
a structured O-mesh is generated in our study. 
The results of the grid convergence study is shown in Fig.~\ref{fig:gc}.
The $C_L/C_D$, when plotted
against the grid factor $N_{cells}^{-2/3}$, exhibits a linear trend, implying second-order convergence.
To balance the calculation accuracy and efficiency, the $L3$ mesh, with dimension $384\times192$ in the wrap-around and normal directions, is used for the current computations.
The computational domain is a 2D mesh with a radius of $~30c$, where $c=1.0$ is the chord length of the airfoil.
By appropriately setting the first layer mesh thickness, the $y+$ of the airfoil surface is less than 1. 
Circumferential grid refinement is performed at the locations where the curvature of the airfoil varies significantly, such as the leading edge and trailing edge. 
The outline and zoomed-in view of the structured mesh adjacent to the airfoil surface are shown in FIG.~\ref{fig:mesh}.

In this study, the compressible Navier-Stokes equation has been employed for simulating the transonic flow around supercritical airfoils, where the effects of body forces and heat sources have been neglected. 
\begin{equation}
\begin{aligned}
\frac{\partial \boldsymbol{U}}{\partial t}+\nabla \cdot \boldsymbol{F}^{\mathrm{c}}-\nabla \cdot\left(\mu_{\mathrm{tot}}^1 \boldsymbol{F}^{\mathrm{v} 1}+\mu_{\mathrm{tot}}^2 \boldsymbol{F}^{\mathrm{v} 2}\right)=0\\
\boldsymbol{U}=\left[\begin{array}{c}\rho \\ \rho \boldsymbol{v} \\ \rho E\end{array}\right],  \boldsymbol{F}^{\mathrm{c}}=\left[\begin{array}{c}\rho \boldsymbol{v} \\ \rho \boldsymbol{v} \otimes \boldsymbol{v}+p \boldsymbol{I} \\ \rho E \boldsymbol{v}+p \boldsymbol{v}\end{array}\right], \boldsymbol{F}^{\mathrm{v} 1}=\left[\begin{array}{c}0 \\ \boldsymbol{\tau} \\ \boldsymbol{\tau} \cdot \boldsymbol{v}\end{array}\right], \boldsymbol{F}^{\mathrm{v} 2}=\left[\begin{array}{c}0 \\ 0 \\ c_p \nabla T\end{array}\right]
\end{aligned}
\end{equation}
Here, $\rho$ is the density of the fluid, $\boldsymbol{v}=\left(v_1, v_2, v_3\right)^{\mathrm{T}}$ is the fluid velocity in the Cartesian reference coordinate system, $E$ is the total energy per unit mass of the fluid, $p$ is the static pressure, $c_p$ is the specific heat at constant pressure, and $T$ is the temperature in Kelvin.
The viscous stress tensor is defined as $\boldsymbol{\tau}=\nabla v+$ $\nabla \boldsymbol{v}^{\mathrm{T}}-\frac{2}{3} \boldsymbol{I}(\nabla \cdot \boldsymbol{v})$.

\begin{figure}[htpb]
	\centering
	\begin{minipage}[t]{0.45\textwidth}
		\centering
		\includegraphics[width=\textwidth]{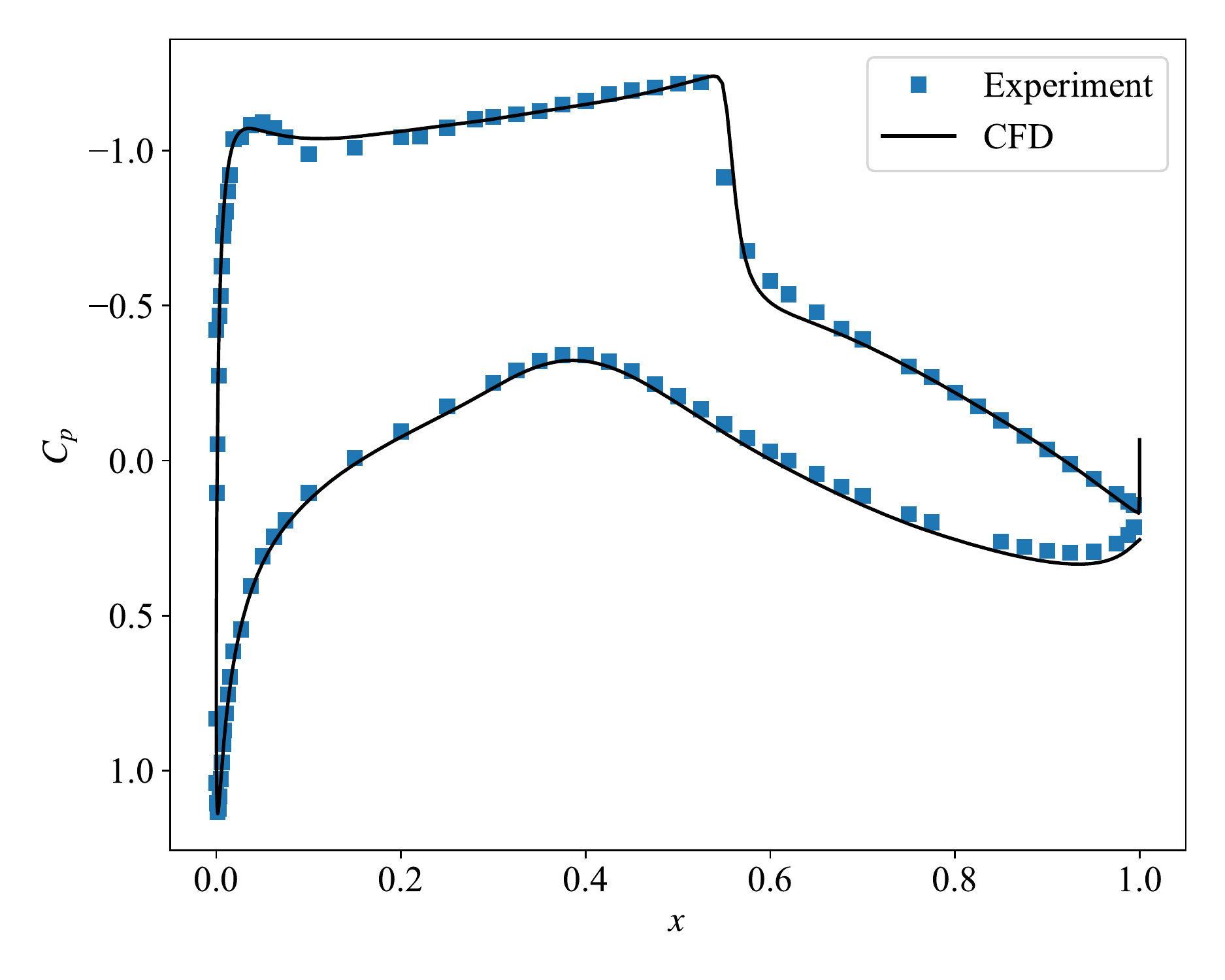}
		\caption{Comparison of pressure coefficient distribution of the RAE2822 with experiment result}
		\label{fig:rae}
	\end{minipage}
	\hspace{10pt}
	\begin{minipage}[t]{0.51\textwidth}
		\centering
		\includegraphics[width=\textwidth]{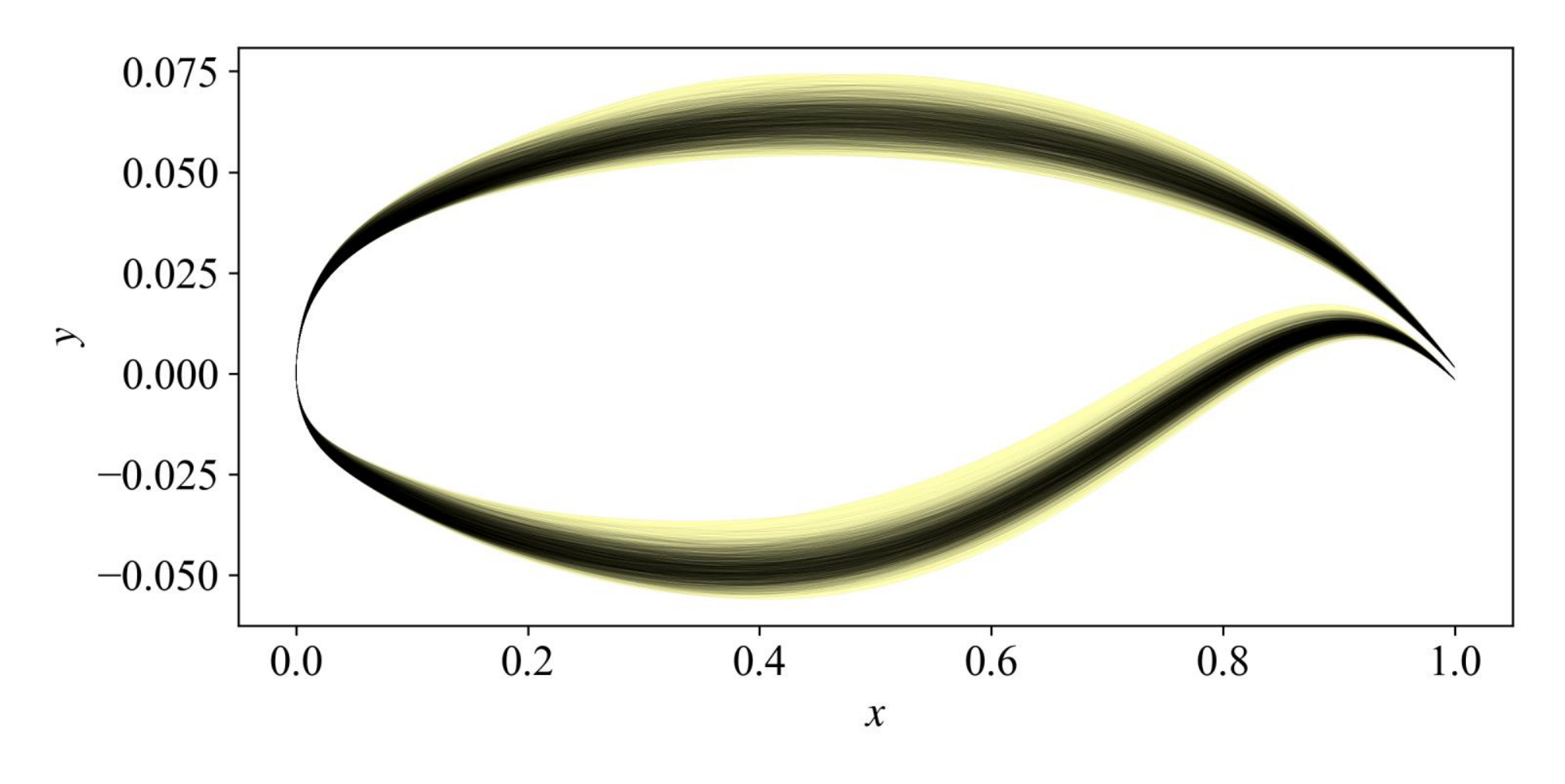}
		\caption{Sampled airfoil geometries}
		\label{fig:airfoil}
	\end{minipage}
\end{figure}

These equations are calculated with the open source code CFL3D~\cite{krist1998cfl3d}, which has been widely employed in engineering applications.
The shear stress transport (SST)~\cite{menter1994assessment} is adopted for turbulence modeling. 
The monotonic upstream-centered (MUSCL) scheme~\cite{van1985upwind} is used to determine state-variable interpolations at the cell interfaces.
The Roe scheme~\cite{roe1981approximate} is used for spatial discretization and the lower-upper symmetric-Gaussian-Seidel (LU-SGS) method~\cite{yoon1988lower} is used for time advancement.
The aerodynamic behaviors of the benchmark RAE2822 airfoil at design flow condition are examined to assess the validity of the numerical scheme. 
FIG.~\ref{fig:rae} shows the comparison of the surface  pressure coefficient ($C_p$) distribution between the CFD simulation and experiment results~\cite{cook1977aerofoil} under the flow condition: $Ma_\infty=0.73, Re=6.5\times10^6, \alpha=3.19^\circ$. 
It shows that the CFD calculation can achieve high accuracy, and the main flow structures including the shock wave can be captured accurately. 

\subsection{Training data}
For subsequent investigations, 500 supercritical airfoils at $Ma=0.73, Re=5.0\times10^6$ under approximately 50 AOAs are simulated by RANS.
Airfoils are parameterized by 14 base functions of Class Shape Transformation (CST)~\cite{kulfan2006fundamental} with the chord length normalized to one.
The dataset used in this study is sampled based on three typical supercritical airfoils, and 
the detailed sampling strategy can refer to the reference \cite{xie2023parametric, runze2022pressure}.
To avoid impractical airfoils and ensure the diversity of the dataset, a set of constraints are implemented during the sampling process. These constraints include enforcing a minimum leading edge radius of 0.007 to avoid impractical geometries. Additionally, the maximum thickness of the airfoil is controlled within a specified range to ensure feasibility. Furthermore, the pressure distribution is regulated to satisfy the flow characteristic of supercritical airfoils. 
By incorporating these constraints, the generated dataset encompasses a wide range of realistic and diverse airfoil geometries suitable for further analysis and design exploration.
The sampled airfoils are depicted in Fig.~\ref{fig:airfoil},
indicating that the dataset involves a wide range of supercritical airfoils.

\begin{figure}[htpb]
	\centering	
	\begin{minipage}[t]{0.47\textwidth}
		\includegraphics[width=\textwidth]{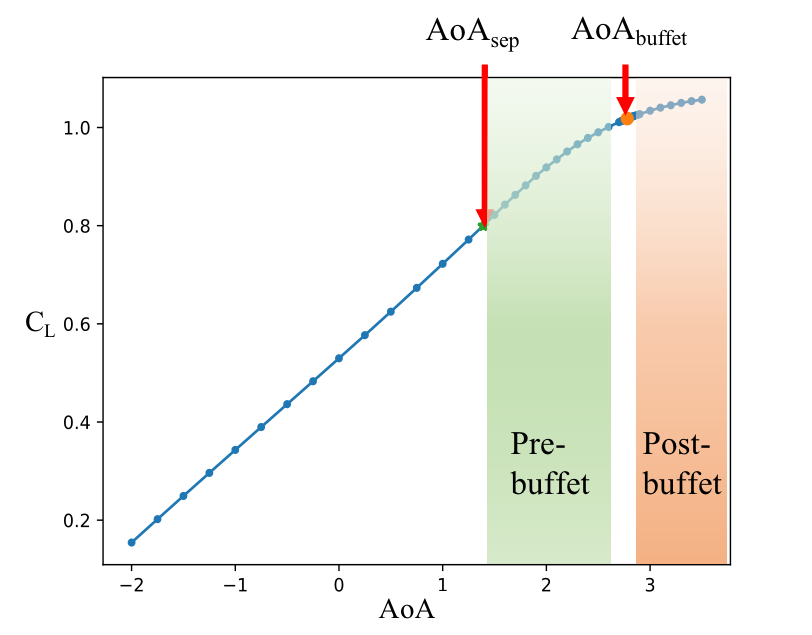}
		\caption{Dataset used for training with buffet onset point predicted by the $\Delta\alpha=0.1^\circ$ method}
		\label{fig:buffet_curve}
	\end{minipage}
	\hspace{16pt}
	\begin{minipage}[t]{0.47\textwidth}
		\centering
		\includegraphics[width=\textwidth]{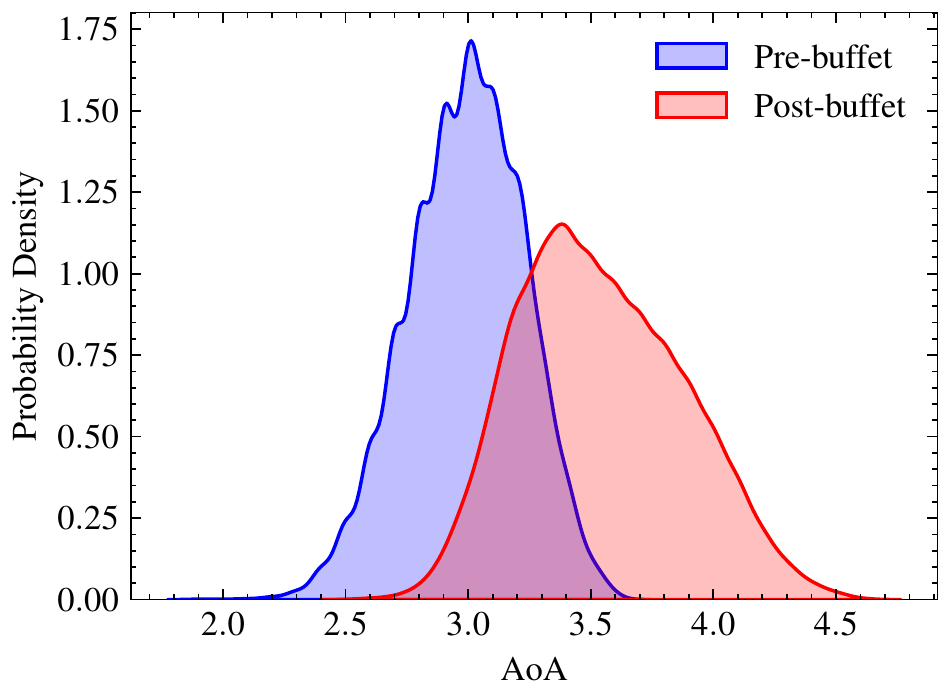}
		\caption{Probability density distributions of AoAs in the training dataset}
		\label{fig:aoa}
	\end{minipage}
\end{figure}

The velocity fields along x-direction, denoted as $u$-velocity, are involved in the training.
The flow is taken as pre-buffet and labeled as 0 if AoA<AoA$_{\text{buffet}}$, 
and is taken as post-buffet and labeled as 1 if AoA>AoA$_{\text{buffet}}$.
Notably, the flow fields without flow separation, i.e., AoA<AoA$_{\text{sep}}$, are excluded from the dataset. 
One reason is to balance the samples of pre- and post-buffet classes;  
the other is to eliminate the influence of unseparated flow to the buffet study.
Besides, given that the ambiguity of the buffet onset predicted from the lift curve, the flow fields at $\text{AoA}_{\text{buffet}} - 0.05^\circ < \text{AoA} < \text{AoA}_{\text{buffet}} + 0.05^\circ$ are also excluded.
Consequently, as depicted in Fig.~\ref{fig:buffet_curve}, the dataset are divided into two classes based on the AoA range to which they are assigned:
\begin{equation}
\begin{aligned}
\text{AoA}_\text{sep} \leq \text{AoA} \leq \text{AoA}_{\text{buffet}} - 0.05^\circ, \quad  \text{ Pre-buffet (0)}\\
\text{AoA} \geq \text{AoA}_{\text{buffet}} + 0.05^\circ, \quad \text{Post-buffet (1)}
\end{aligned}
\end{equation}
Eventually, a total of 26912 samples are generated with 21530 randomly selected samples for training and the remaning 5382 samples for validation.
Fig.~\ref{fig:aoa} presents the probability density distributions of AoAs involved in the training, and it is seen that the dataset involves a wide range of AoAs evenly distributed in pre- and post-buffet conditions.

\section{Theoretical Methods}
The objective of this study is to develop a method to learn a low-dimensional latent space, which is flexible and expressive in capturing the extremely high-dimensional flow field data~\cite{eivazi2022towards,kang2022physics}, while also well-constrained to significantly associated with the transonic buffet.

\subsection{Variational Autoencoder}
Variational Autoencoder (VAE) is a graphical generative model based on the concept of auto-encoding~\cite{kingma2013auto,qu2021deep,yang2022flowfield}.
Consider a sample set $\mathbf{x}\in\mathbb{R}^n$, which are independent and identically distributed random variables with distribution as $p(\mathbf{x})$. 
Assume that the data $\mathbf{x}$ are generated by some random process involving unobserved continuous random variable $\mathbf{z} \in \mathbb{R}^m(m \ll n)$, which are denoted as latent variables.
Given the prior distribution $p(\mathbf{z})$, $\mathbf{x}$ can be sampled from the conditional distribution $p(\mathbf{x}|\mathbf{z})$, which is usually achieved through a decoder network with $\theta$ as learned parameters.
\begin{equation}
\mathbf{x}' =\text { Decoder}(\mathbf{z};\mathbf{\theta})
\end{equation}
Similarly, to infer the latent variables $\mathbf{z}$ from the input data $\mathbf{x}$, 
an encoder is usually adopted to model the distribution $p(\mathbf{z}|\mathbf{x})$.
Due to the intractability of the posterior $p(\mathbf{z}|\mathbf{x})$, 
variational inference is adopted to find an approximate probability distribution $q_\phi(\mathbf{z}|\mathbf{x})$ with $\phi$ as learned parameters as close as possible to $p(\mathbf{z}|\mathbf{x})$. 
Then, the model minimizes the difference between $q_\phi(\mathbf{z}|\mathbf{x})$ and $p(\mathbf{z}|\mathbf{x})$, which is described by the Kullback-Leibler divergence (KLD) as follows:
\begin{equation}
\begin{aligned}
\mathrm{KL}\left(q_\phi(\mathbf{z}|\mathbf{x}) \| p(\mathbf{z}|\mathbf{x})\right) & =\int_z q_\phi(\mathbf{z}|\mathbf{x}) \log \frac{q_\phi(\mathbf{z}|\mathbf{x})}{p(\mathbf{z}|\mathbf{x})}d\mathbf{z}  \\
&=\int_\mathbf{z} q_\phi(\mathbf{z}|\mathbf{x}) \log \frac{p(\mathbf{x}, \mathbf{z})}{q_\phi(\mathbf{z}| x)}d\mathbf{z}+\log p(\mathbf{x})
\end{aligned}
\end{equation}

Since $p(\mathbf{x})$ is independent of $\phi$, the problem turns into minimizing the first term $\mathcal{L}(\phi, \theta)$ and it can be expressed as:
\begin{equation}\label{eq:Lphi}
\mathcal{L}(\phi, \theta)=
\int_\mathbf{z} q_\phi(\mathbf{z}|\mathbf{x}) \log \frac{p(\mathbf{x}, \mathbf{z})}{q_\phi(\mathbf{z}| x)}d\mathbf{z}=
\underbrace{\mathbb{E}_{q_\phi(\mathbf{z} | \mathbf{x})}\left[\log p_\theta(\mathbf{x} | \mathbf{z})\right]}_{\mathcal{L}_{Recon}}
+\underbrace{\mathrm{KL}\left(q_\phi(\mathbf{z}|\mathbf{x})\| p(\mathbf{z})\right)}_{\mathcal{L}_{KLD}}
\end{equation}
The first term $\mathbb{E}_{q_\phi(\mathbf{z} | \mathbf{x})}\left[\log p_\theta(\mathbf{x} | \mathbf{z})\right]$ denotes the reconstruction error between the ground truth data $\mathbf{x}$ and the generated data $\mathbf{x}'$, 
and it can be calculated by the following expression:
\begin{equation}\label{eq:mse_loss}
\mathcal{L}_{Recon}= \frac{1}{N}\sum_{i=1}^{N}(\frac{1}{2}||\mathbf{x}'_{i}-\mathbf{x}_{i}||^{2})
\end{equation}
The second term $\mathrm{KL}\left(q_\phi(\mathbf{z}|\mathbf{x})\| p(\mathbf{z})\right)$ measures the difference between $q_\phi(\mathbf{z}|\mathbf{x})$ and $p(\mathbf{z})$, and encourages the approximate posterior to be close to the prior $p(\mathbf{z})$. 
However,  the gradient with respect to $\phi$ is difficult to estimate
since $\phi$ appears in the distribution with respect to which the expectation is taken.
Therefore, the multivariate Gaussian distribution $q_\phi(\mathbf{z}| \mathbf{x})$ is
reparameterized using a noise distribution $p(\epsilon)$ and a differentiable transformation $g_\phi$. 
Assume $\epsilon$ are sampled from a multivariate unit Gaussian, i.e., $p(\epsilon) \sim \mathcal{N}(0, \mathbf{I})$. 
And if we let 
\begin{equation}
\mathbf{z}=g_\phi(\epsilon, \mathbf{z})=\mu_\phi(\mathbf{x})+\epsilon \odot \sigma_\phi(\mathbf{x})
\end{equation}
where $\odot$ denotes elementwise multiplication. $\mathbf{z}$ will have the desired distribution $q_\phi(\mathbf{z} | \mathbf{x}) \sim \mathcal{N}\left(\mathbf{z} ; \mu_\phi(\mathbf{x}), \sigma_\phi(\mathbf{x})\right)$. 
Then, the encoder is parameterized with $\mathbf{\phi}$ and expressed as
\begin{equation}
\begin{aligned}
(\mathbf{\mu}, \log \mathbf{\sigma}) =\text { Encoder}(\mathbf{x};\mathbf{\phi}) \\
q_\phi(\mathbf{z} | \mathbf{x}) =\mathcal{N}(\mathbf{z} ; \mathbf{\mu}, \operatorname{diag}(\mathbf{\sigma}))
\end{aligned}
\end{equation}

Accordingly, the second term in Eq.\eqref{eq:Lphi} can be rewritten as below when the posterior $q_\phi(\mathbf{z} | \mathbf{x})$ and prior $p_\theta(\mathbf{z})$ are assumed to follow the Gaussian distribution $N\left(\mu, \sigma^2\right)$ and $N\left(0, \mathbf{I}^2\right)$, respectively.
\begin{equation}
\mathcal{L}_{KLD}=\frac{1}{2} \sum_{k=1}^n\left(\sigma_k^2+\mu_k^2-\left(\log \left(\sigma_k^2\right)+1\right)\right)
\end{equation}
where $\mu_k$ and $\sigma_k$ represent the mean and standard deviation of $z_k$. 
As the posterior distribution $q_\phi(\mathbf{z} | \mathbf{x})$ approaches the prior distribution $p_\theta(\mathbf{z})$, the KL-divergence decreases.

Finally, the loss function of the VAE model with $\beta$ balancing the reconstruction error and KL-divergence~\cite{burgess2018understanding}, can be formulated as: 
\begin{equation}
\begin{aligned}
\mathcal{L}_{VAE}= \mathcal{L}_{Recon}+\beta\mathcal{L}_{KLD}
\end{aligned}
\end{equation}

\subsection{Physics-Assisted Variational Autoencoder (PAVAE)}
The latent space $\mathbf{z}$ in VAE is a low-dimensional representation of high-dimensional system dynamics.
In many situations, the latent space is also expected to be well-constrained so as to highly correlated with other characteristics.
Suppose a set of categorical labels of $\mathbf{x}$ is $\mathbf{c}\in\{0,1\}$, where 0 denotes the pre-buffet state and 1 denotes the post-buffet state in our case.
The latent space $\mathbf{z}$ is fed into a classifier, which outputs the corresponding classification output $\mathbf{y}\in\mathbb{R}$. 
From another point of view, if the classifier is incorporated into the encoder of VAE,
a composite classifier is formed with $\mathbf{x}$ as input and  $\mathbf{y}$ as output.
To better understand the criteria of classification, the classifier in our study is designed to be the simplest form with one layer as follows:
\begin{equation}\label{eq:class}
\mathbf{y} = \sigma\left(\mathbf{W}\mathbf{z}+b\right)
= \sigma\left(\mathbf{W}\left(\mu_\phi(\mathbf{x})+\epsilon \odot \sigma_\phi(\mathbf{x})\right)+b\right)
\end{equation}
where $W: \mathbb{R}^{m} \rightarrow \mathbb{R}$ is a linear transformation and $b\in\mathbb{R}$ is a bias term. $\sigma: \mathbb{R} \rightarrow \mathbb{R}$ is a non-linear activation function. The Sigmoid function
$\sigma(x)=1 /\left(1+e^{-x}\right)$ is chosen in this study, which maps the output into $[0,1]$ range that can be interpreted as a probability.

The binary cross-entropy (BCE) loss is used to train the network, assuming independent and identically distributed training examples:
\begin{equation}
\begin{aligned}
\mathcal{L}_{B C E} & =-\frac{1}{N} \sum_{i=1}^N c_i\left(\ln y_i\right)+\left(1-y_i\right)\left(\ln \left(1-c_i\right)\right)
\end{aligned}
\end{equation}
where $c_i$ and $y_i$ denote the true binary label and the predicted probability of the $i$th sample.

Take the binary cross-entropy loss into the VAE network, and the PAVAE network is then trained by minimizing the following total loss:
\begin{equation}
\begin{aligned}
\mathcal{L}_{PAVAE}=\mathcal{L}_{Recon}+\beta\mathcal{L}_{KLD}+\alpha\mathcal{L}_{B C E}
\end{aligned}
\end{equation}
Where $\alpha$ is the weight to adjust the contribution of the classifier to the whole network.
In addition, mean absolute error (MAE) that describes the mean difference between the reconstructed and the original field,  and maximum absolute error (MAXE) that describes the maximum difference between the reconstructed and the original field, are also used as indicators to evaluate the reconstruction performance of the model.
The overall procedure of proposed method is summarized in
Algorithm~\ref{alg:pavae}.

\begin{algorithm}[H]
	\caption{Physics-Assisted Variational Autoencoder} 
	\label{alg:pavae}
	\begin{algorithmic}[1]
		\STATE Generate dataset $\mathbf{x}$ and corresponding labels $\mathbf{c}$.\
		\FOR{$epoch=1, \ldots, N_{\text {epoch}}$}
		\FOR{$k=1, \ldots, N_{\text {mini-batch}}$}
		\STATE 
		$\mathbf{\mu}_i, \mathbf{\sigma}_i =\text { Encoder}(\mathbf{x}_i), \quad i=1, \ldots, N_k$ ($N_k$ is the number of samples in $k$th mini-batch);
		\STATE
		$\mathcal{L}_{KLD} = \frac{1}{2} \sum_{i=1}^{N_k}\left(\sigma_i^2+\mu_i^2-\left(\log \left(\sigma_i^2\right)+1\right)\right)$;
		\STATE 
		$\mathbf{z}_i \sim \mathbf{\mu}_i+\mathbf{\sigma}_i \odot \epsilon, {\epsilon} \sim N(\mathbf{0}, \mathbf{I})$;
		\STATE 
		${\mathbf{x}'}_i=\text{Decoder}\left(\mathbf{z}_i\right)$;
		\STATE 
		$\mathcal{L}_{recon} = \frac{1}{N_k} \sum_{i=1}^{N_k}\left(\mathbf{x}_i-{\mathbf{x}}'_i\right)^2$;
		\STATE	 ${\mathbf{y}}_i=\text{Classifier}\left(\mathbf{z}_i\right)$;		
		\STATE
		$\mathcal{L}_{BCE} = -\frac{1}{N_k} \sum_{i=1}^{N_k} \mathbf{c}_i\left(\ln \mathbf{y}_i\right)+\left(1-\mathbf{y}_i\right)\left(\ln \left(1-\mathbf{c}_i\right)\right)$;
		\STATE
		$loss = \mathcal{L}_{Recon}+\beta\mathcal{L}_{KLD}+\alpha\mathcal{L}_{B C E}$;
		\STATE
		Update model with the Adam optimizer;
		\ENDFOR
		\ENDFOR
	\end{algorithmic}
\end{algorithm}

\subsection{Detailed network}
\begin{figure}[htpb]
	\centering
	\includegraphics[width=\linewidth]{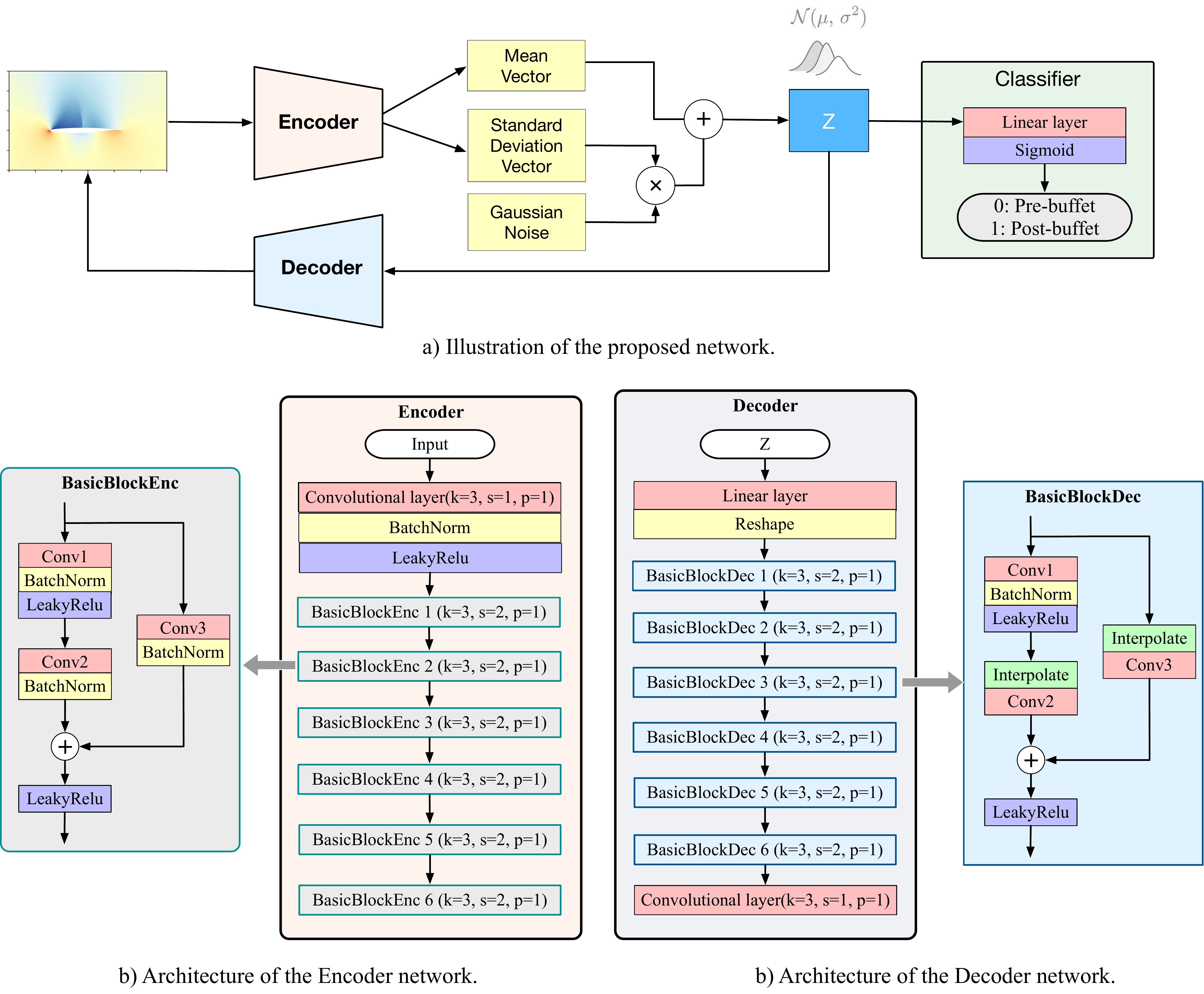}
	\caption{Architecture of the detailed network}
	\label{fig:network}
\end{figure}

As we know, Convolutional Neural Network (CNN) is good at efficiently handling the spatial information of input images to approximate the required output through highly nonlinear mapping. 
Therefore, we implement the encoder and decoder in our model as a deep CNN, so as to efficiently represent the spatio structure of systems governed by partial differential equations~\cite{wu2017introduction,han2019novel}.
FIG.~\ref{fig:network} depicts the detailed network used in our study, which is simplified for clarity purposes. 

The raw input of the network is the u-velocity flow field, 
whose format is similar to the processing method of pictures and videos in computer vision task. 
The physical variable of each grid in the computation domain, space direction, and the number of physical variables are taken as a pixel, height/width size, and RGB channel of an image, respectively.
Thus, the input in our case is expressed as a {three-dimensional (3D) matrix of size $1\times384\times192$.

The encoder is designed to include six convolutional residual blocks, similar to that in ResNet~\cite{szegedy2017inception}. 
The utilization of the residual connection can strengthen the depth of the network, which can accelerate the convergence speed of the deep network without introducing any redundant parameters and increasing computational complexity.
Each of the convolutional residual blocks follows the same pattern, which consists of a convolutional layer, a batch normalization layer, an activation function, and a residual connection.
3$\times$3 convolution layer is performed with a fixed feature map channel $[16, 32, 64, 128, 256]$, respectively, bypassing the input every two convolutions. 
The leaky rectified linear unit (LeakyRelu) with a slope of 0.2 is employed as the activation function. 
According to these operators in blocks, the high-resolution fields can be reduced with increasing channels, and complicated nonlinear mapping is achieved simultaneously.
Following several convolutional residual blocks, the mean and standard deviation for each input can be extracted. Thus a compact features $\mathbf{z}$ of the input data is transmitted to the decoder.

The decoder is also designed to comprise six convolutional residual blocks. 
The up-sampling process is utilized to expand the input size gradually, aiming to realize the transformation from low-dimensional data to high-dimensional data.
An upsample layer replaces the convolution layer of the encoder in the residual block, including a bilinear interpolation filter to interpolate small-sized features without weights and a convolution layer to interpret the interpolated inputs into meaningful details.
A kernel of size 3, stride 1, and padding 1 is used across all over the convolution layers.
The feature map dimension is $[256, 128, 64, 32, 16]$ respectively, bypassing the input every two convolutions.

The dimensionality of the latent space $\mathbf{z}$ is important for the performance of the models.
Too small dimension may lead to insufficient reconstruction. 
Too large dimension may lead to overfitting and is unfavourable for interpretation.
Thereby, the technique based on POD proposed in our previous study~\cite{wang2021flow} is adopted to attain a rational latent dimension.
The basic idea is that VAE typically requires fewer dimensions than POD for equivalent reconstruction accuracy, which has been discussed and proven in many studies~\cite{agostini2020exploration,murata2020nonlinear}.
Thus, POD, which takes little cost and effort, can be regarded as a crude accuracy indicator for the same dimension of VAE.
According to the explained variance ratios of different principal components obtained from our dataset, which are presented in Fig.~\ref{fig:pca}, the latent dimension is determined to be 10 since it conserves 98.4\% in terms of energy contents of POD, which is judged to be sufficient.
\begin{figure}[htpb]
	\centering
	\includegraphics[width=\textwidth]{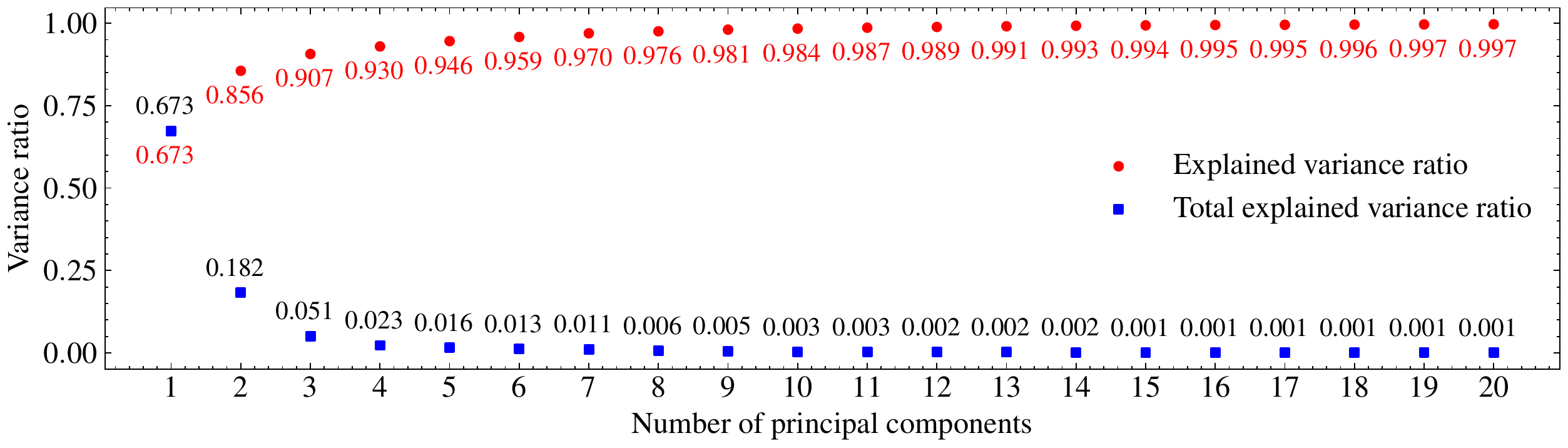}
	\caption{Explained variance ratios of first 20 principal components.}
	\label{fig:pca}
\end{figure}

The implementation, training, and testing for the PAVAE based models are performed using PyTorch.  
The detailed hyperparameter settings for the framework are described as follows. 
Four models with $\alpha\in [0.0, 1.0, 10.0, 100.0]$ are trained, so as to investigate the influence of additional physical information to the latent space. 
It is easy to derive that the PAVAE degenerates into a VAE when $\alpha=0.0$.
The flow fields' data are normalized to $[0, 1]$ range through a pre-processing step. 
A fixed batch size of 16 is applied.
The loss function is optimized using the Adam optimizer~\cite{kingma2014adam} and the weight $\beta$ is setting to be $1\times{10}^{-4}$.
The Cosine learning rate warmup is employed~\cite{gotmare2018closer}, where the learning rate increases linearly from 0 to $0.005$ over 100 epochs, and
decreases from $0.005$ to 0 over remaining 900 epochs following a cosine curve.

\section{Results and discussions}

\subsection{Training Results}
The convergence histories of the total loss, reconstruction loss, and BCE loss obtained with different $\alpha$ are illustrated in Fig.~\ref{fig:loss}. 
The detailed convergent errors are summarized in Table~\ref{tab:loss}, along with a comparison of the results obtained from the POD. 
It should be noted that the classifier with $\alpha=0$ fails to update due to zero gradient, resulting in  $\mathcal{L}_{BCE}$ that does not accurately reflect the classification performance of the latent space. 
To address this issue, the classifier with $\alpha=0$ is retrained with the encoder and decoder parameters frozen after the completion of the training process.
The proposed models achieve complete convergence of all errors after 1,000 epochs, with both the MAE and MAXE converging to magnitudes of $10^{-4}$ and $10^{-2}$, respectively. 
These magnitudes are one order of magnitude higher than those achieved by POD.
The accuracy of the proposed models reaches almost 100\%, demonstrating their classification ability. 
Notably, even without guidance from an additional classifier, $\alpha=0$ achieves high accuracy due to the latent space's ability to capture the primary characteristics associated with buffet.

\begin{figure}[htpb]
	\centering
	\includegraphics[width=\textwidth]{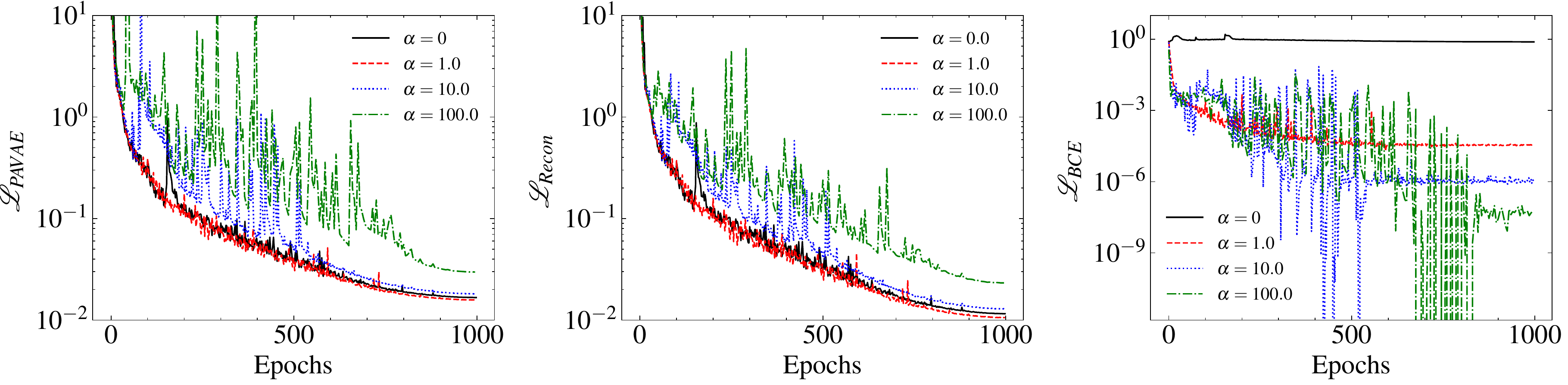}
	\caption{Convergence histories of four models with varying $\alpha$ values, depicting the total loss, reconstruction loss, and BCE loss trajectories in separate subfigures.
 }
	\label{fig:loss}
\end{figure}

\begin{table}[htpb]
	\caption{Comparison of convergence errors of four models using varying $\alpha$}
	\label{tab:loss}%
	\centering
        \begin{ruledtabular}
	\begin{tabular}{@{}cccccc@{}}
		& MAE  & MAXE	& $\mathcal{L}_{KLD}$	&$\mathcal{L}_{BCE}$	&Accuracy\\
		\hline
        POD & 0.003978 & 0.551580 & - & 0.001293 & 1.000 \\
		$\alpha=0.0$&	0.000224& 0.040604&	51.233865&	0.006183 &	0.999\\
		$\alpha=1.0$&	0.000213& 0.038828&	51.553975&	0.000036&	1.000\\
		$\alpha=10.0$&	0.000246& 0.040530&	52.106018&	0.000001&	1.000\\
		$\alpha=100.0$&	0.000347& 0.041237&	64.530717&	0.000000&	1.000\\
	\end{tabular}
        \end{ruledtabular}
\end{table}

As $\alpha$ increases, the classification performance of the proposed method improves, with the best performance achieved at $\alpha=100.0$ as indicated by the lowest BCE loss and highest accuracy.
This suggests that the proposed method promotes the proximity of the latent space to the physical information, specifically the buffet state in this study. 
However, this improvement in classification performance comes at the expense of a decrease in reconstruction performance. 
For example, the value of $\alpha=100.0$ results in the worst reconstruction performance with the maximum values of MAE, MAXE, and KLD loss. 
The discrepancy in the latent space between the reconstruction and classification tasks is responsible for this trade-off, as increasing $\alpha$ brings a set of latent variables closer to the classification latent space, while pushing them further away from the reconstruction latent space. 
Nonetheless, a smaller value of $\alpha$, such as $\alpha=1.0$, leads to better performance in both classification and reconstruction tasks compared to $\alpha=0.0$, due to the overlapping latent space between the two tasks. The high classification accuracy obtained using the latent space derived from the reconstruction task suggests a connection between the two latent spaces. 
Overall, an appropriate $\alpha$ can be associated with an appropriate intrinsic physics, resulting in an optimal latent space with good performance in both reconstruction and classification tasks.

Figure~\ref{fig:recon} depicts the reconstructed u-velocity fields and velocity profiles on the upper side of the airfoil at $x/c=0.8$ for three randomly selected samples, providing a comprehensive demonstration of the proposed models' flow field reconstructive ability. 
The reconstructed contour patterns obtained using the four models demonstrate a satisfactory match with the CFD results. 
Specifically, coarse structures are identically reproduced, and fine structures near the shock wave are closely approximated with minor differences. 
Additionally, the velocity profiles are accurately reconstructed and match identically with the ground truth data. 
The four models with varying $\alpha$ values do not exhibit any significant differences in their ability to handle and compress high-dimensional turbulent flow data, indicating a superior performance in terms of flow field reconstruction.

\begin{figure}[htpb]
	\centering
	\hspace{25pt}
		\includegraphics[width=0.3\linewidth]{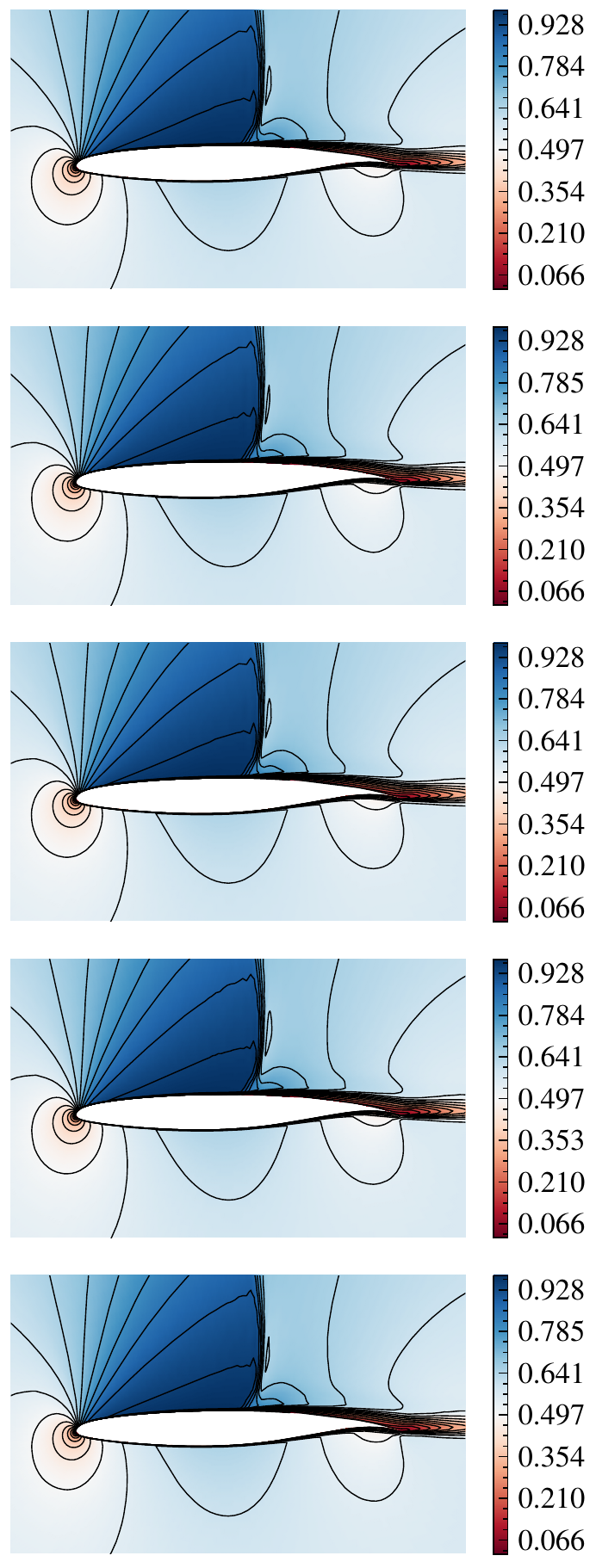}
		\put(-180,330){CFD}
		\put(-180,260){$\alpha=0.0$}
		\put(-180,180){$\alpha=1.0$}
		\put(-180,110){$\alpha=10.0$}
		\put(-185,35){$\alpha=100.0$}
		\includegraphics[width=0.3\linewidth]{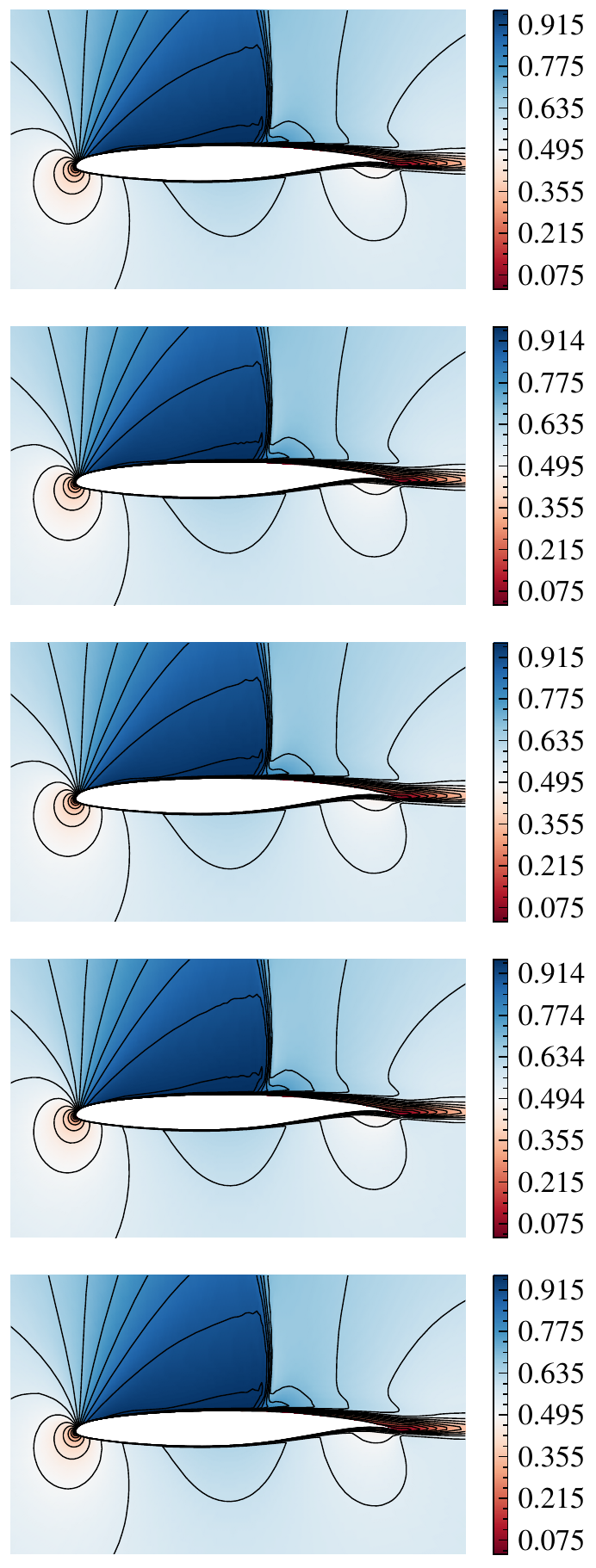}
		\includegraphics[width=0.3\linewidth]{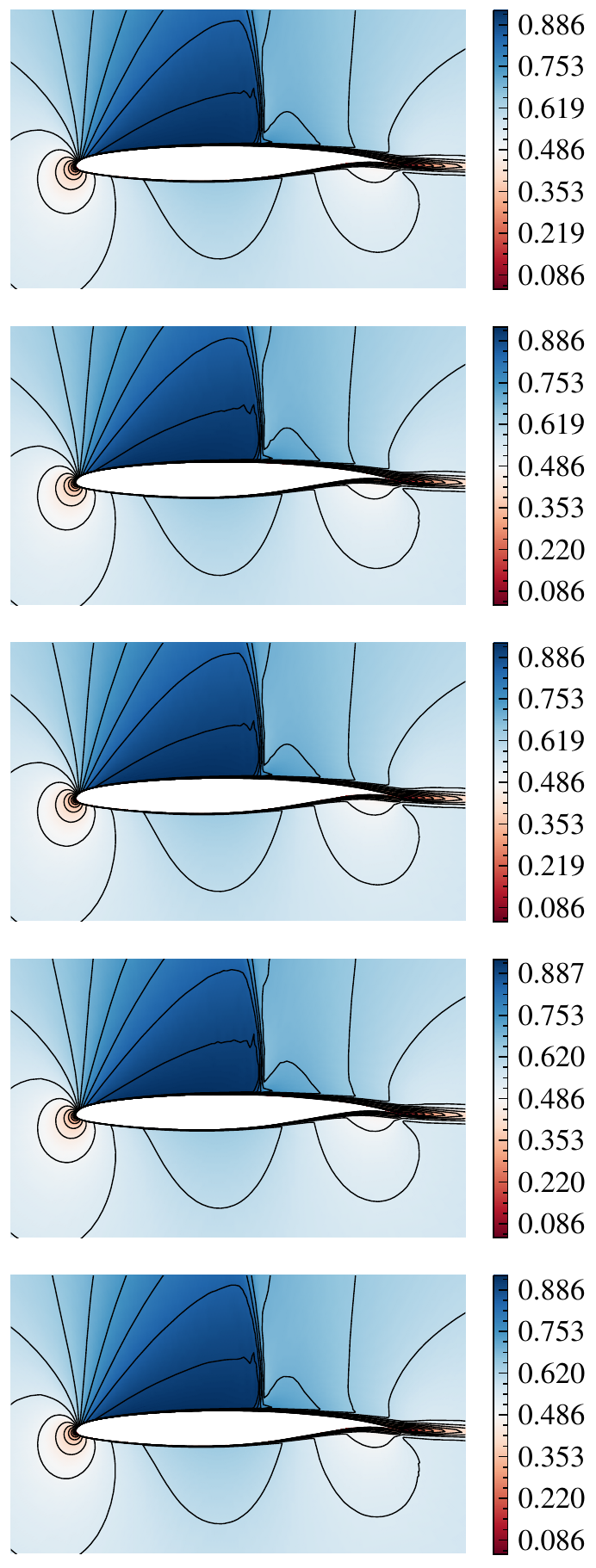}
	\vfill
	\hspace{18pt}
	\subfigure[Sample 1]{
		\includegraphics[width=0.3\linewidth]{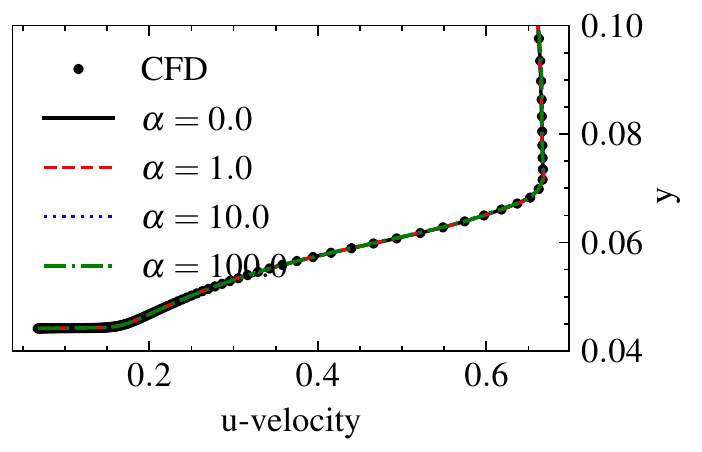}
		\put(-180,50){Velocity}
		\put(-177,35){Profile}
	}
	\subfigure[Sample 2]{
		\includegraphics[width=0.3\linewidth]{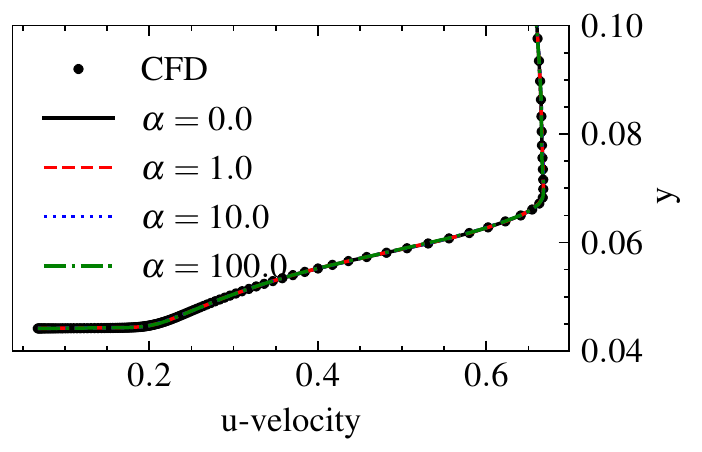}
	}
	\subfigure[Sample 3]{
		\includegraphics[width=0.3\linewidth]{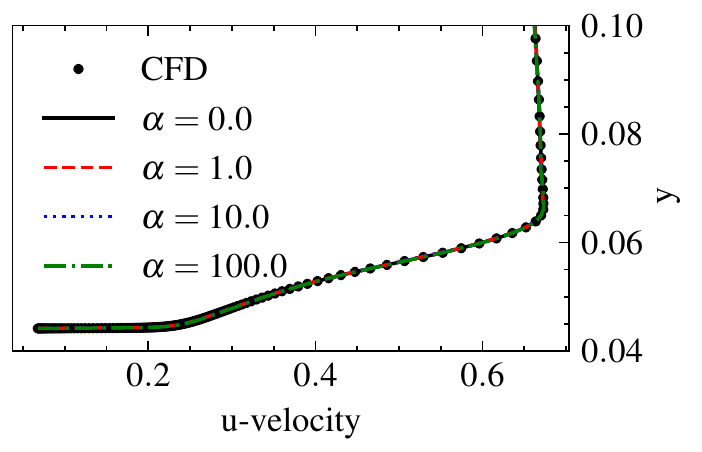}
	}
	\caption{Reconstructed u-velocity fields and velocity profiles on the upper side of the airfoil at $x/c=0.8$ of three randomly selected samples}
	\label{fig:recon}
\end{figure}

To assess the impact of $\alpha$ on the latent space, the Uniform Manifold Approximation and Projection (UMAP) method~\cite{mcinnes2018umap} was employed to visualize the 10-dimensional latent representations in a two-dimensional (2D) space. 
The UMAP technique aims to learn the manifold structure of high-dimensional data and identify a low-dimensional embedding that preserves the essential topological structure of the manifold. 
Fig.~\ref{fig:umap} illustrates the UMAP projection of the four models, colored by the buffet labels. 
The results suggest that the incorporation of additional physical information can significantly impact the latent space. 
As expected, clustering by individuals is more visible in the latent space obtained with $\alpha=10.0$ and $\alpha=100.0$ than in those obtained with $\alpha=0.0$ and $\alpha=1.0$. 
The latent representations are perfectly grouped and separable across individuals when $\alpha=100.0$. 
Hence, it can be concluded that larger $\alpha$ leads to a latent space that can capture more significant differences between pre- and post-buffet samples.

\begin{figure}[htpb]
	\centering
	\includegraphics[width=\textwidth]{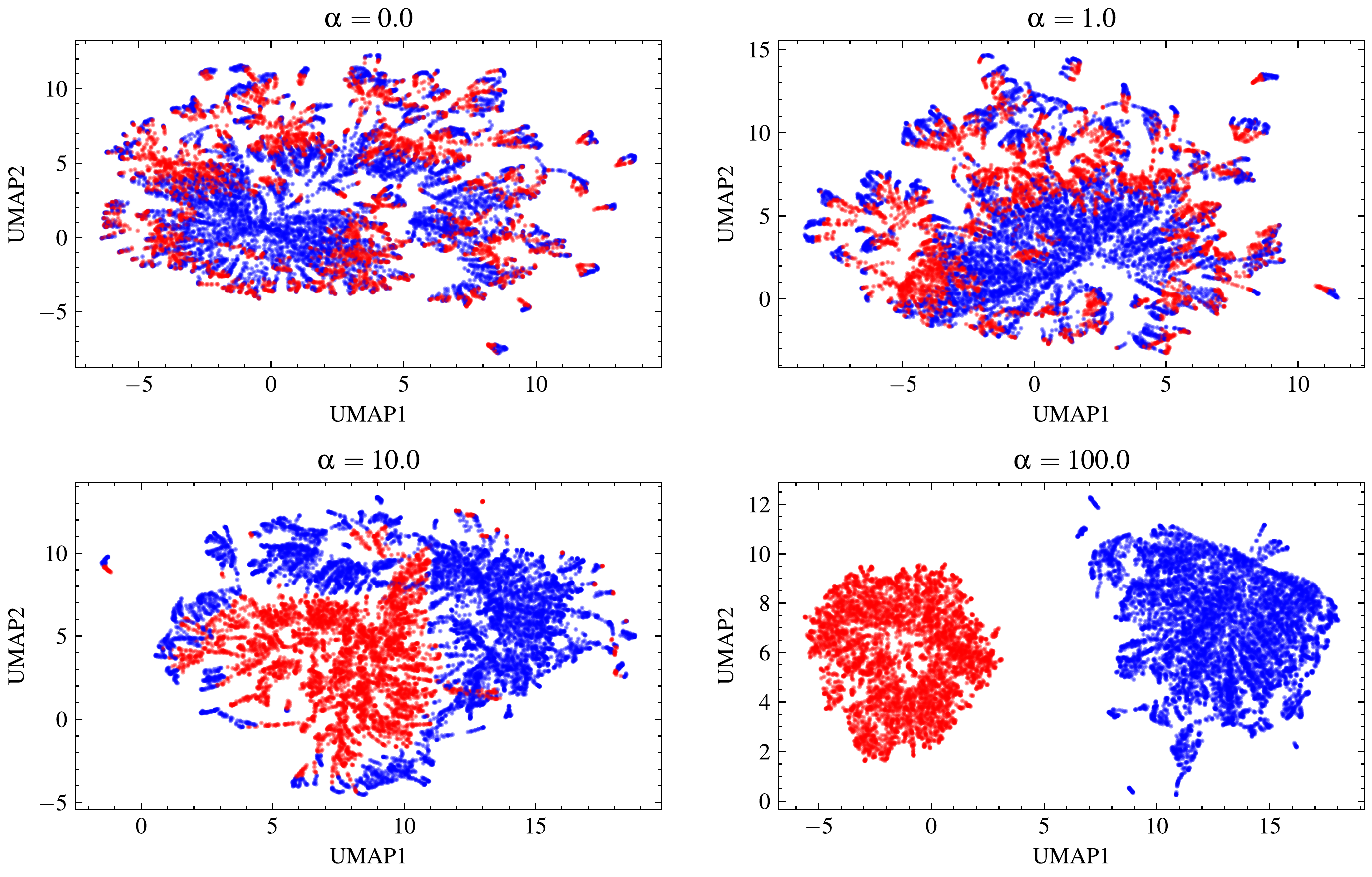}
	\caption{UMAP projections of the latent spaces obtained with different $\alpha$ colored by buffet labels (Blue: Pre-buffet, Red: Post-buffet)}
	\label{fig:umap}
\end{figure}

\subsection{Identifying dominant regions of transonic buffet}

The successful implementation of the proposed models indicates that the classifiers can effectively discriminate between pre- and post-buffet flows, and the latent space can comprehensively characterize the transonic buffet. 
To explore the contribution of each latent space to transonic buffet, $\mathbf{W}$ and statistics of $\mathbf{W}\mathbf{z}$ in Eq.~\ref{eq:class} are plotted in Fig.~\ref{fig:contribution}.
Here, $\mathbf{W}$ denotes the overall contribution of each latent space across all samples, while $\mathbf{W}\mathbf{z}$ represents the specific contribution of each latent space to each sample.
Larger $\mathbf{W}$ and $\mathbf{W}\mathbf{z}$ indicate that the corresponding latent space has larger contribution to the classifier and stronger relevance to the transonic buffet.
Conversely, close-to-zero $\mathbf{W}$ and $\mathbf{W}\mathbf{z}$ values indicate that the corresponding latent variable is less important and its contribution to the classifier is negligible. 
Consequently, the ten latent variables are ranked based on their $\mathbf{W}$ and $\mathbf{W}\mathbf{z}$ values, and we observe that most of the latent space can be regarded as inactive and has little relevance to the transonic buffet. Hence, the first three dominant latent spaces in each model are retained for further investigation.

\begin{figure}[htpb]
	\centering
	\includegraphics[width=\linewidth]{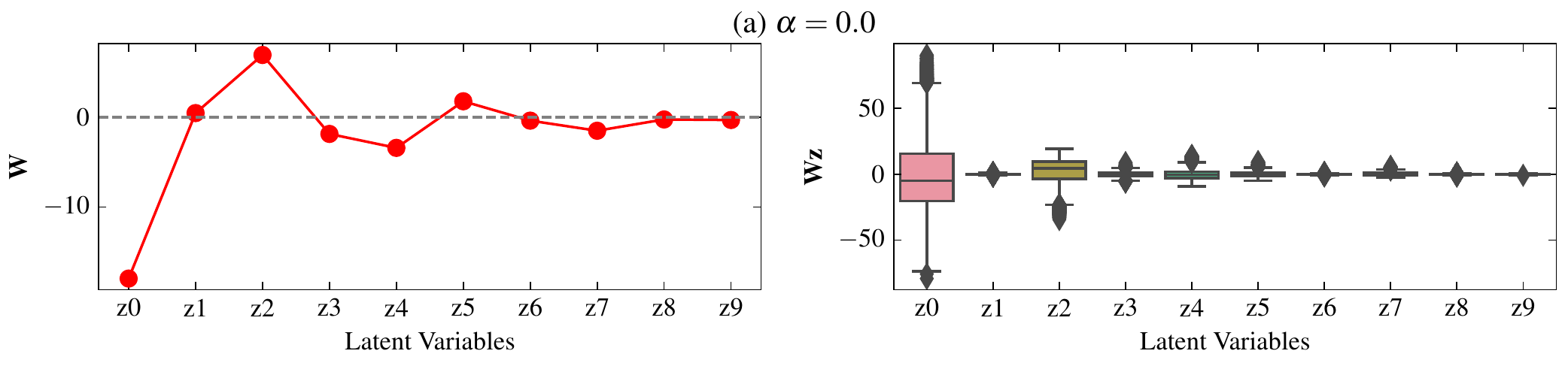}
	\includegraphics[width=\linewidth]{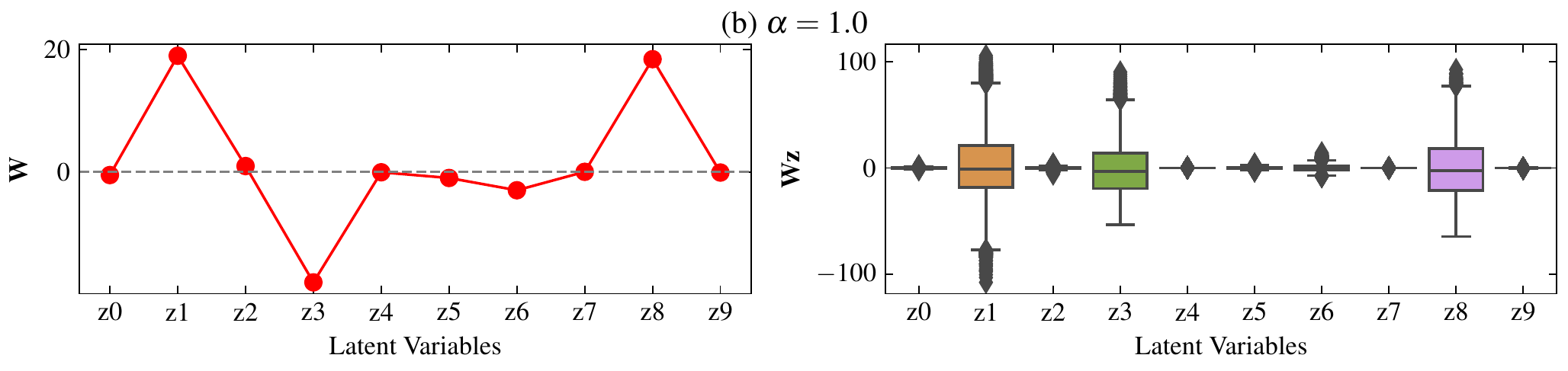}
	\includegraphics[width=\linewidth]{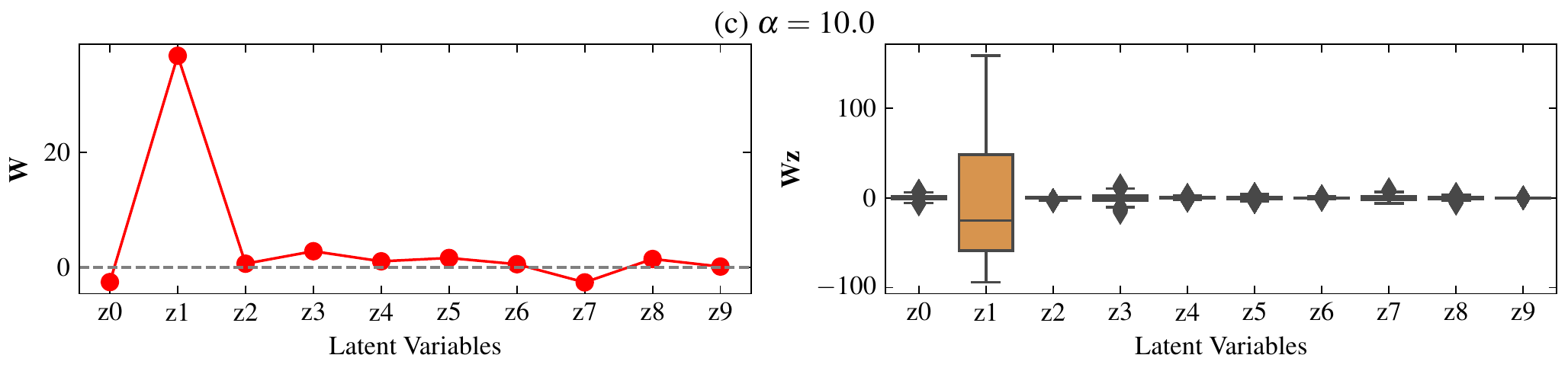}
	\includegraphics[width=\linewidth]{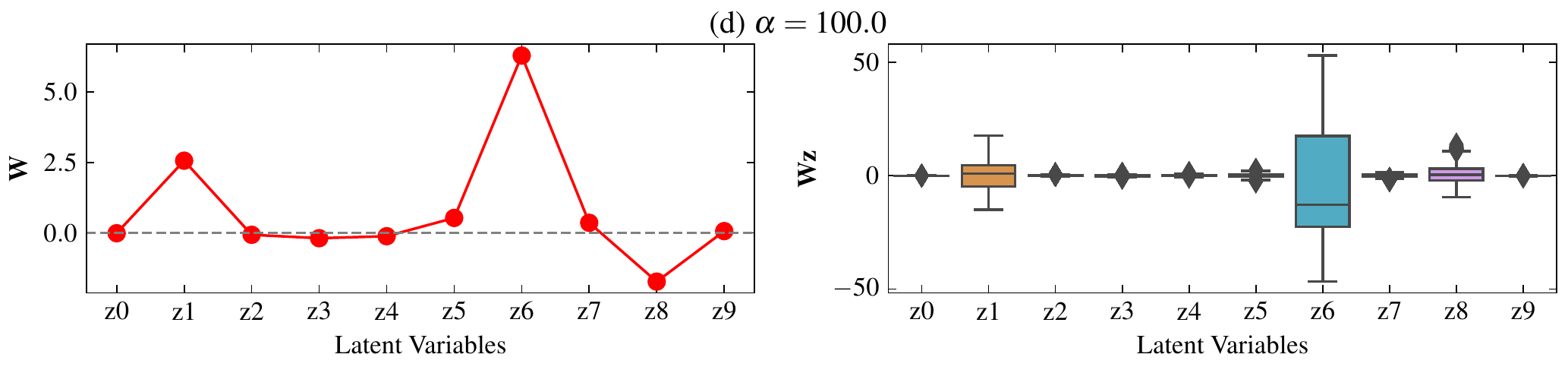}
	\caption{$\mathbf{W}$ and statistics of the $\mathbf{W}\mathbf{z}$ for each latent space}
	\label{fig:contribution}
\end{figure}

To investigate the relationship between the dominant latent space and transonic buffet, logistic regression is employed to classify the buffet using various dimensional latent space, whose results are listed in Table.~\ref{tab:class}.
When using the 3D latent space, constructed by the first three dominant latent variables, all four models achieve nearly 100\% accuracy. 
The 3D latent space obtained from $\alpha=0.0$ and $\alpha=1.0$ are shown in Fig.~\ref{fig:latentspace}(a)-(b), colored by the buffet labels.
The pre- and post-buffet samples exhibit a certain level of differentiation. 
Notably, the distinction between pre- and post-buffet samples is less clear for $\alpha=0.0$ compared to $\alpha=1.0$, corresponding to the lower classification accuracy of the former's latent space. 
When using the 2D latent space, constructed by the first two dominant latent variables, $\alpha=0.0$ and $\alpha=1.0$ exhibit limited classification ability, with much lower accuracy than the 3D latent space.
In contrast, $\alpha=10.0$ and $\alpha=100.0$ achieve 100\% accuracy with a clear boundary, as shown in Fig.~\ref{fig:latentspace}(a)-(b).
Interestingly, it is found that $\alpha=10.0$ and $\alpha=100.0$ achieve nearly complete differentiation using  one dimensional (1D) latent space constructed by the most dominant latent variable.
These findings suggest that the transonic buffet state can be accurately determined using only one dominant characteristic.

\begin{figure}[htpb]
	\centering
	\subfigure[$\alpha=0.0$]
	{
		\begin{minipage}[t]{0.45\textwidth}
			\includegraphics[width=1\linewidth]{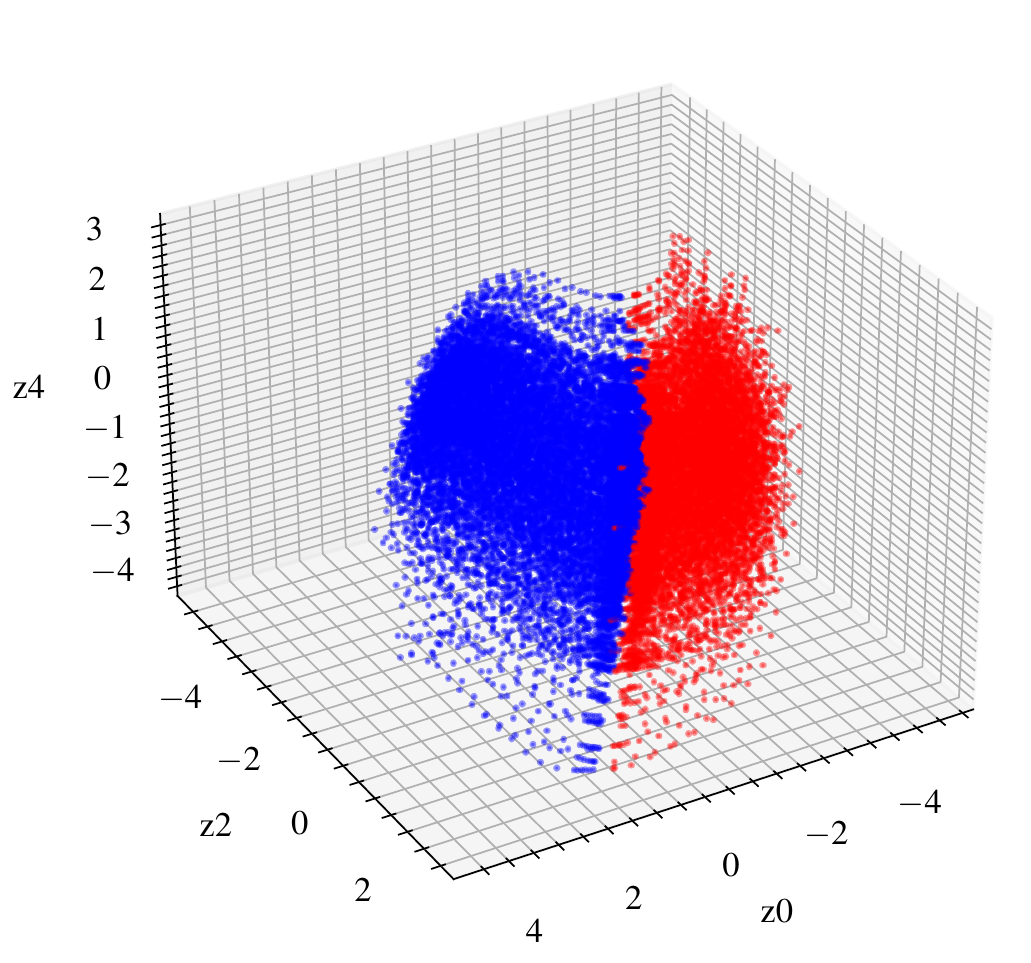}
		\end{minipage}
	}
	\hspace{10pt}
	\subfigure[$\alpha=1.0$]
	{
		\begin{minipage}[t]{0.45\textwidth}
			\includegraphics[width=1\linewidth]{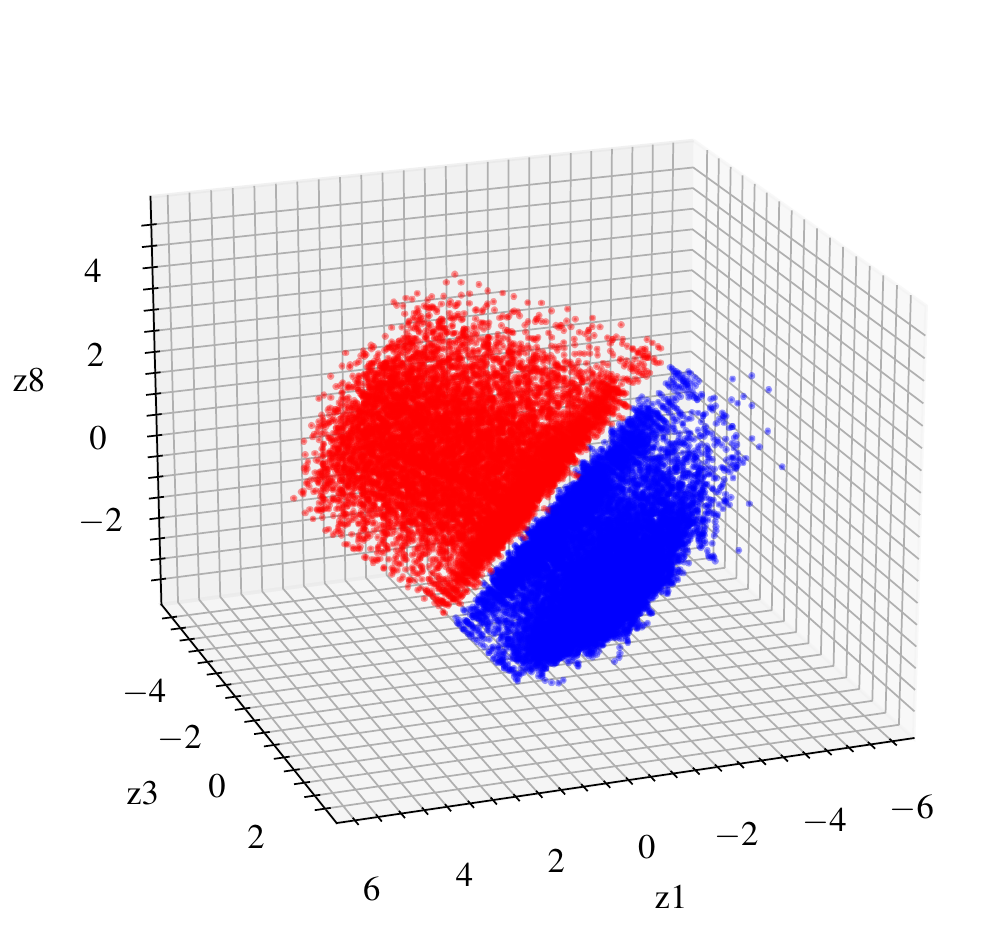}	
		\end{minipage}
	}
	\subfigure[$\alpha=10.0$]
	{
		\begin{minipage}[t]{0.45\textwidth}
			\includegraphics[width=1\linewidth]{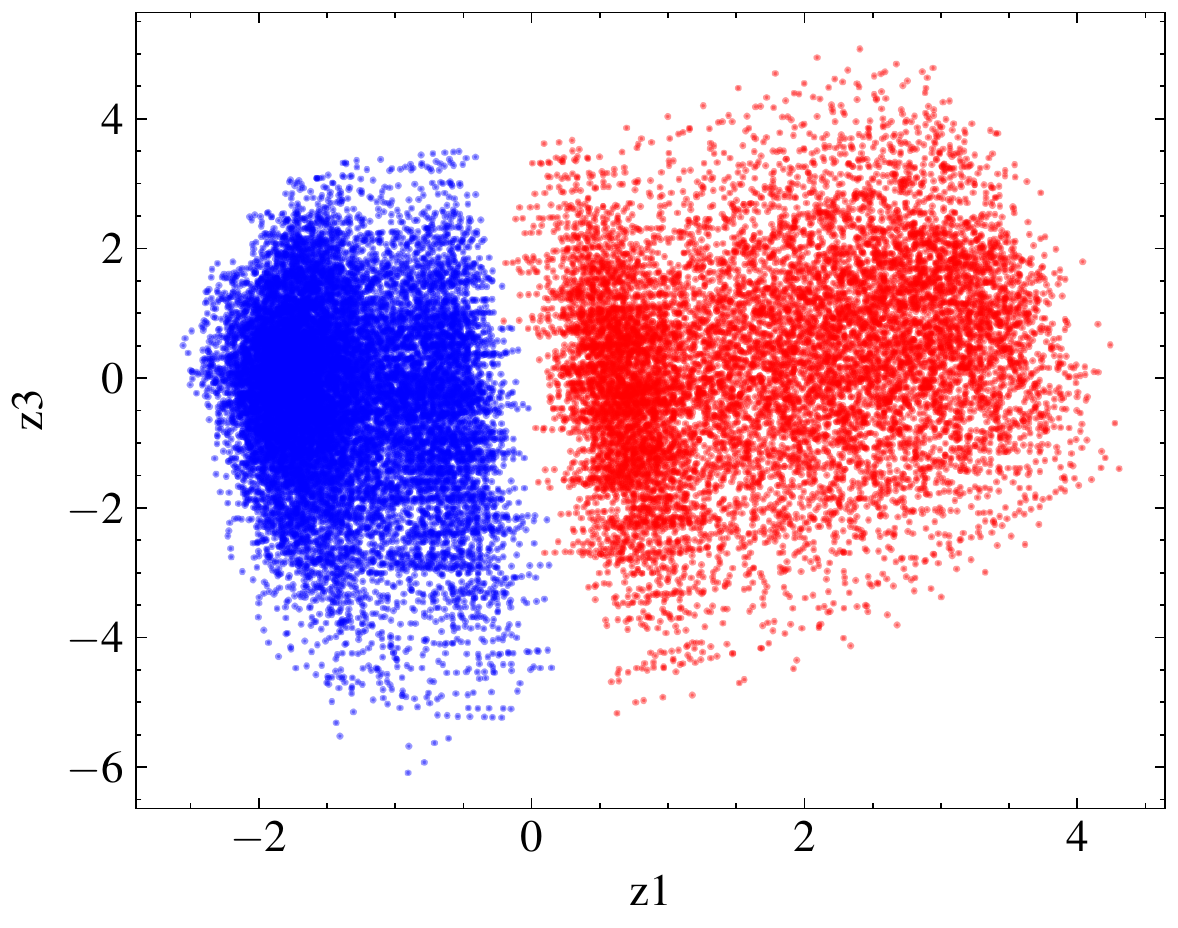}	
		\end{minipage}
	}
	\hspace{10pt}
	\subfigure[$\alpha=100.0$]
	{
		\begin{minipage}[t]{0.45\textwidth}
			\includegraphics[width=1\linewidth]{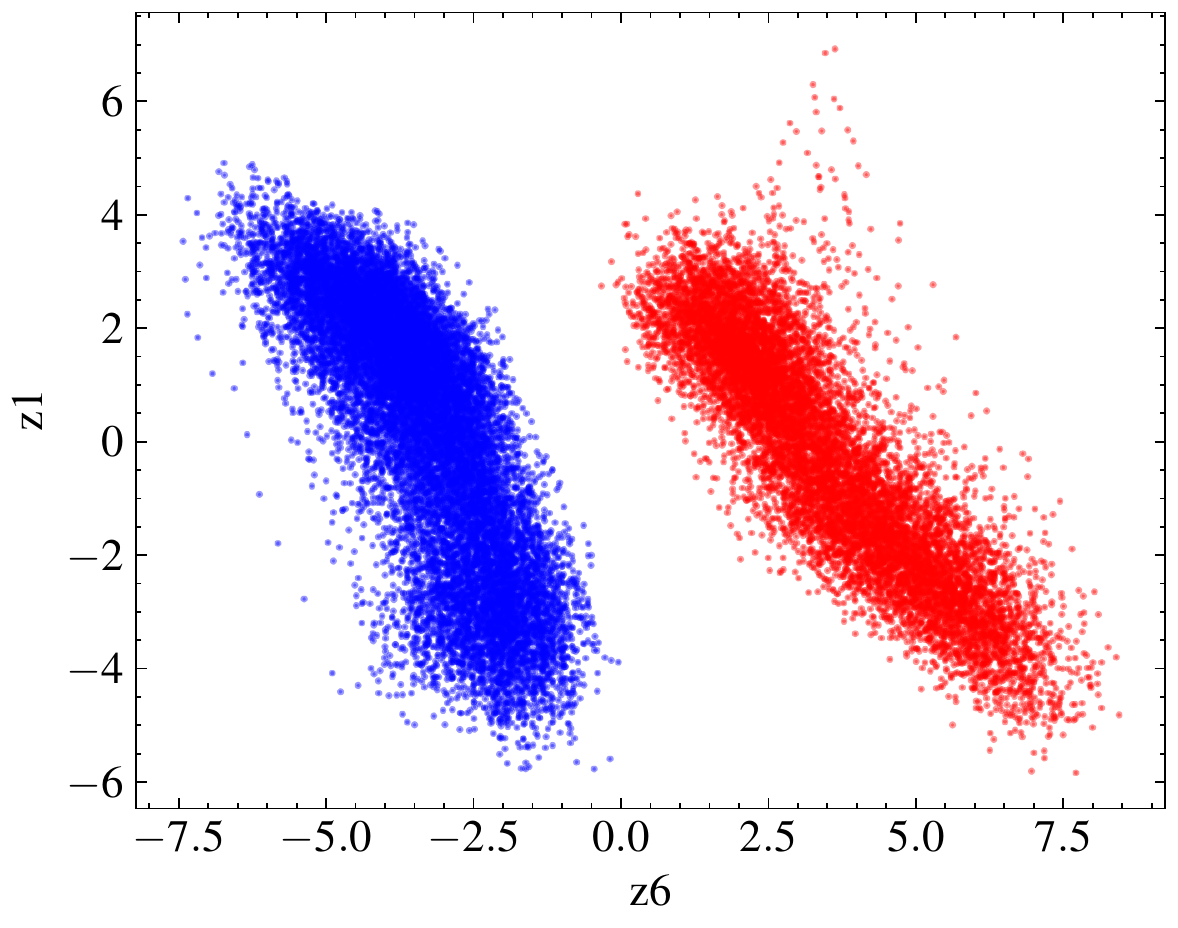}	
		\end{minipage}
	}
	\caption{Dominant latent spaces obtained with different $\alpha$ colored by the buffet labels (Blue: pre-buffet, Red: post-buffet)
	}
	\label{fig:latentspace}	
\end{figure}

\begin{table}[htpb]
	\caption{Classification results using various dimensional latent space}
	\label{tab:class}%
	\centering
	\begin{ruledtabular}
	\begin{tabular}{@{}ccccccc@{}}
		\toprule
		&\multicolumn{2}{c}{1D latent space} &\multicolumn{2}{c}{2D latent space} &	\multicolumn{2}{c}{3D latent space}\\
		\cmidrule{2-7}
		&Latent Variables&	Accuracy&	Latent Variables&	Accuracy&	Latent Variables&	Accuracy\\
		\hline
		$\alpha=0.0$&	z0&	0.884&	z0, z2&	0.977&	z0, z2, z4&	0.992\\
		$\alpha=1.0$&	z1&	0.779&	z1, z3&	0.906&	z1, z3, z8&	
		0.999\\
		$\alpha=10.0$&	z1&	0.999&	z1, z3&	1.000&	z1, z3, z7&	
		1.000\\
		$\alpha=100.0$&	z6&	1.000&	z6, z1&	1.000&	z6, z1, z8&	1.000\\
	\end{tabular}
        \end{ruledtabular}
\end{table}

\begin{figure}[htpb]
    \centering
	\subfigure[$\alpha=0.0$]{
		\centering
		\includegraphics[width=0.33\linewidth]{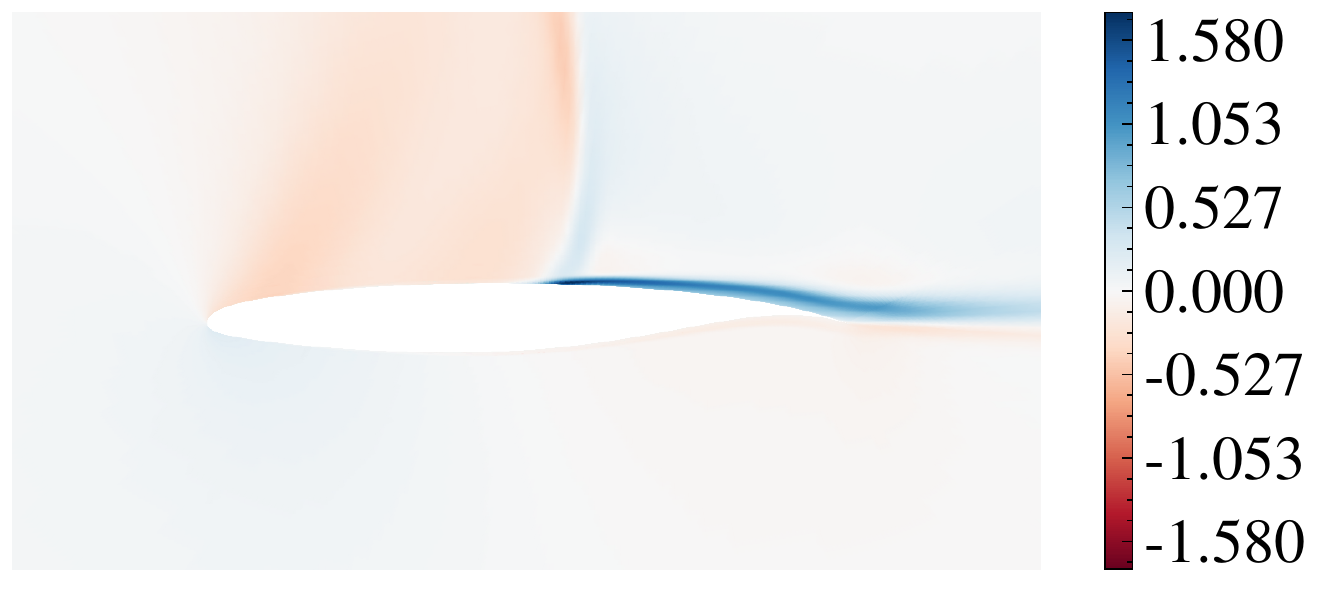}
            \put(-150,60){\textbf{z0}}
		\includegraphics[width=0.33\linewidth]{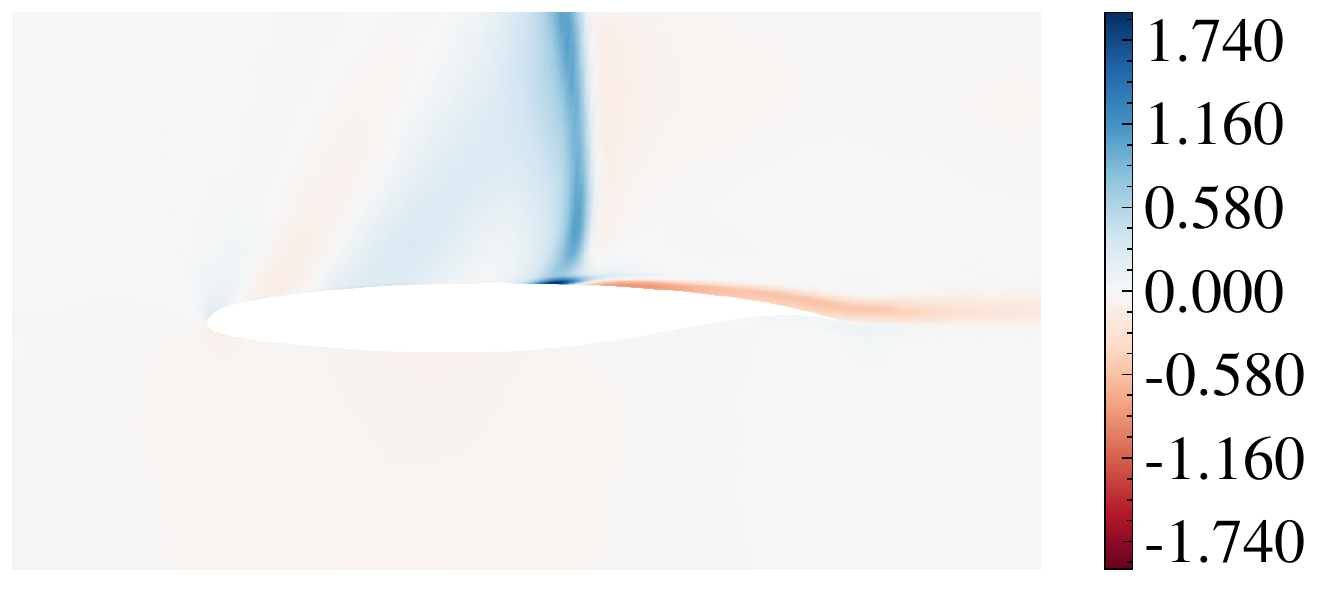}
            \put(-150,60){\textbf{z2}}
		\includegraphics[width=0.33\linewidth]{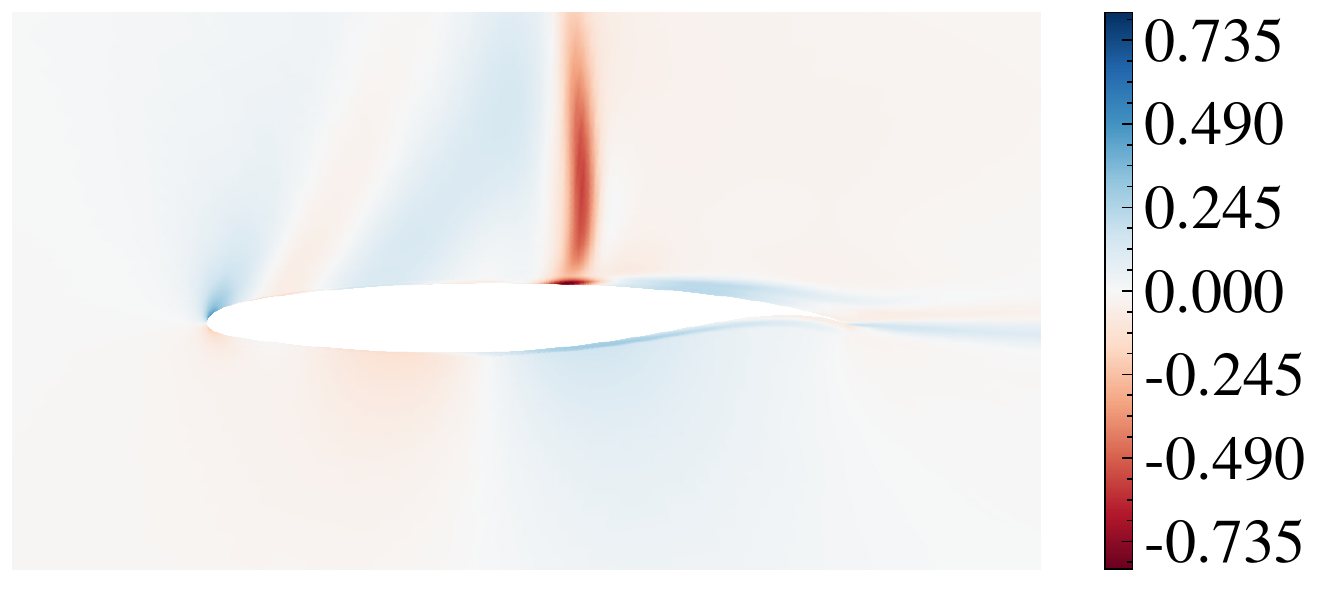}
            \put(-150,60){\textbf{z4}}
        }
	\subfigure[$\alpha=1.0$]{
		\centering
		\includegraphics[width=0.33\linewidth]{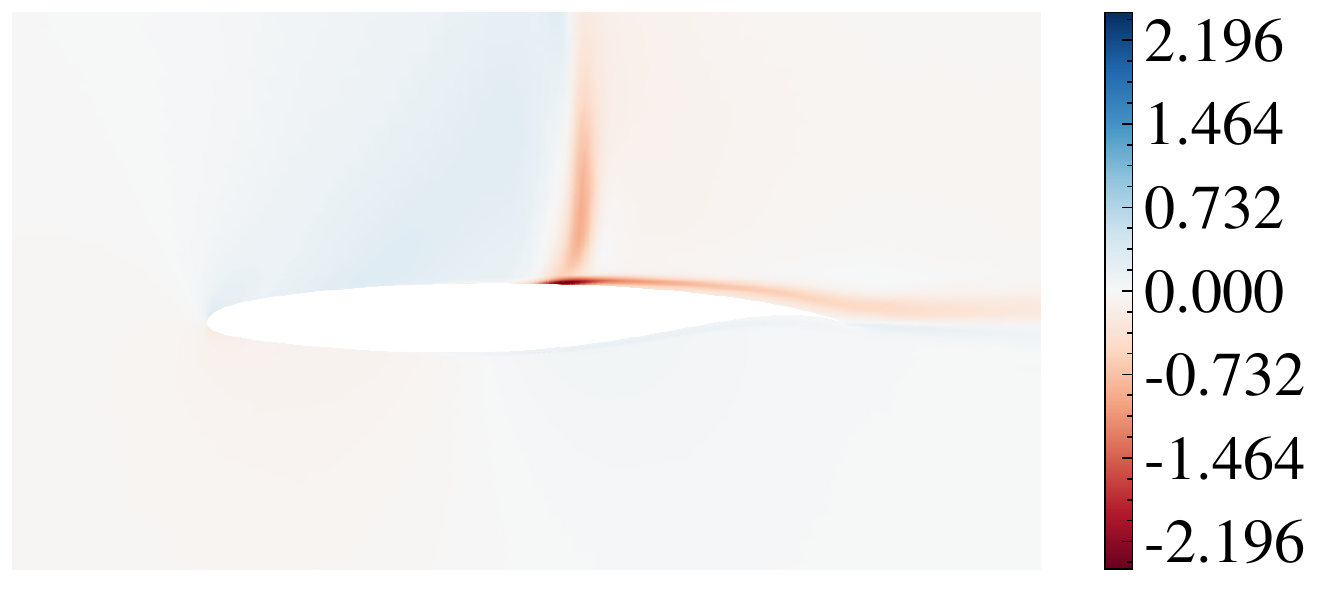}
            \put(-150,60){\textbf{z1}}
		\includegraphics[width=0.33\linewidth]{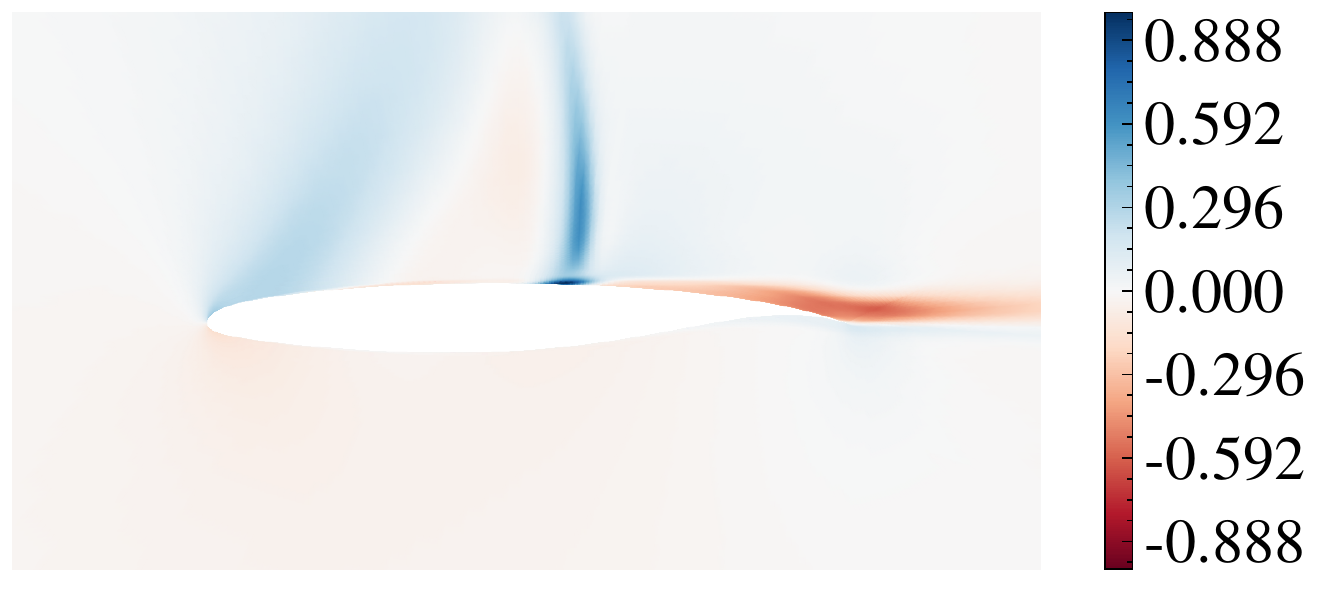}
            \put(-150,60){\textbf{z3}}
		\includegraphics[width=0.33\linewidth]{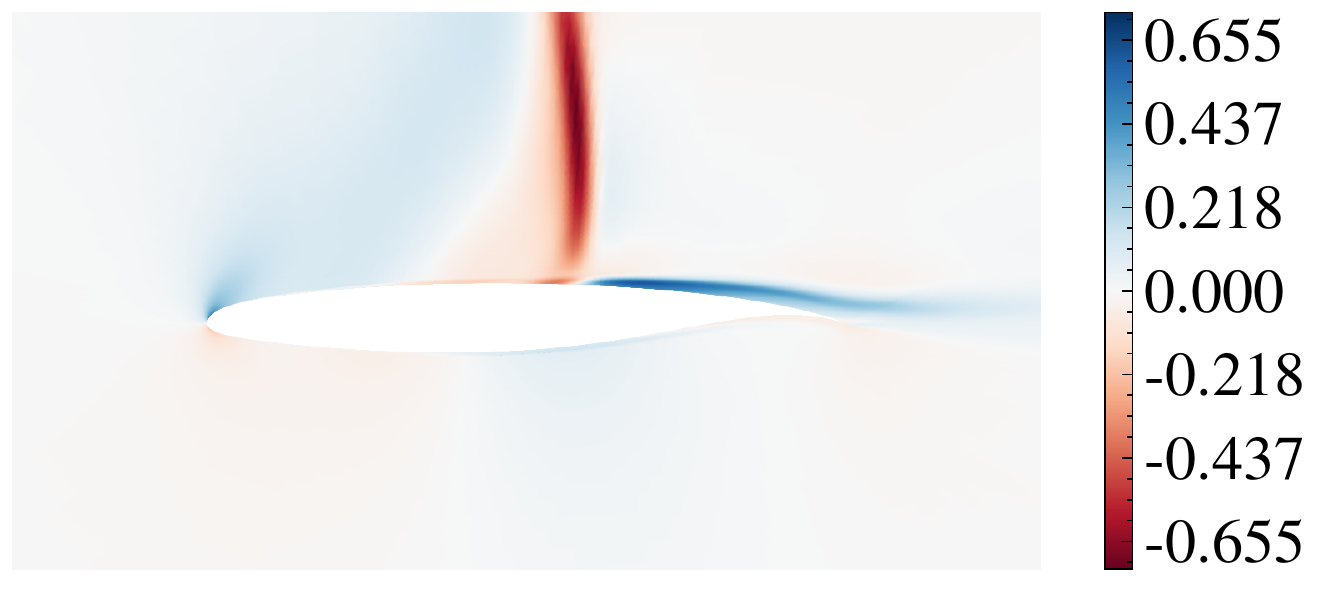}
            \put(-150,60){\textbf{z8}}
        }
	\subfigure[$\alpha=10.0$]{
		\centering
		\includegraphics[width=0.33\linewidth]{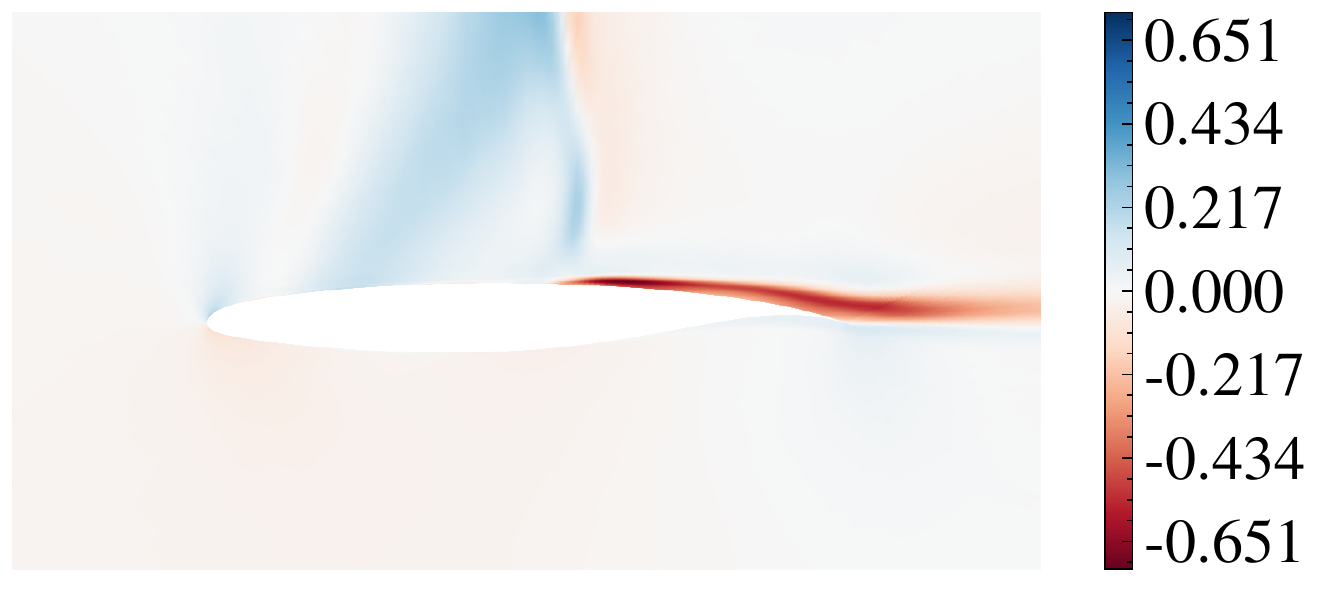}
            \put(-150,60){\textbf{z1}}
		\includegraphics[width=0.33\linewidth]{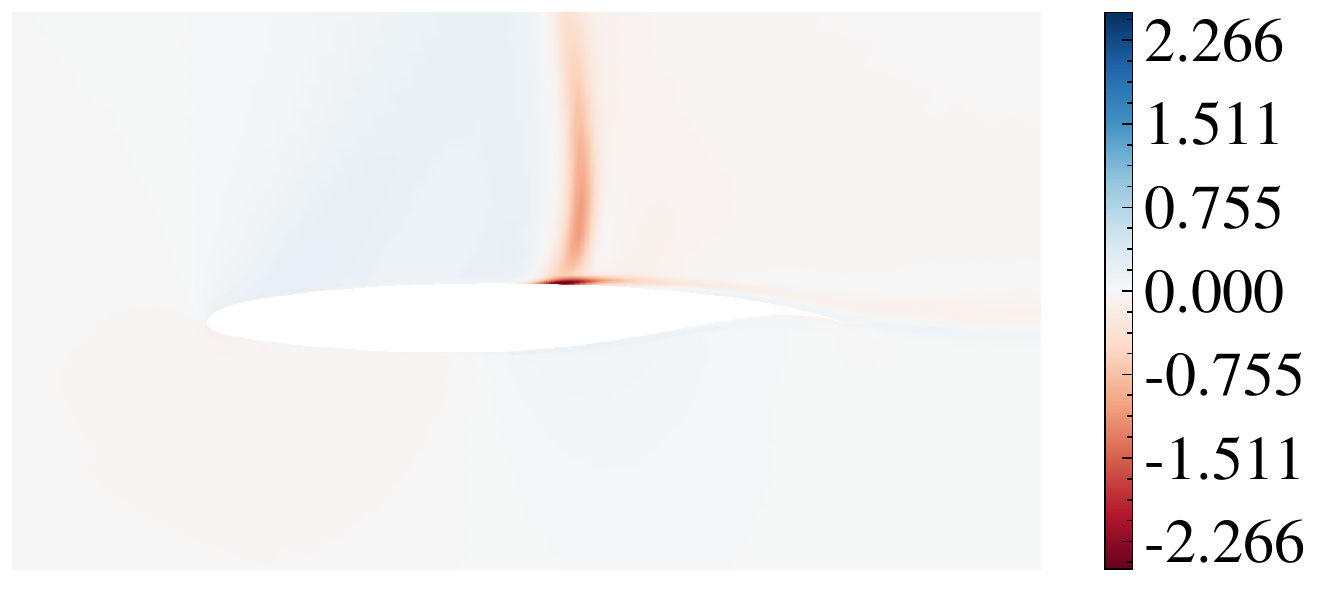}
            \put(-150,60){\textbf{z3}}
		\includegraphics[width=0.33\linewidth]{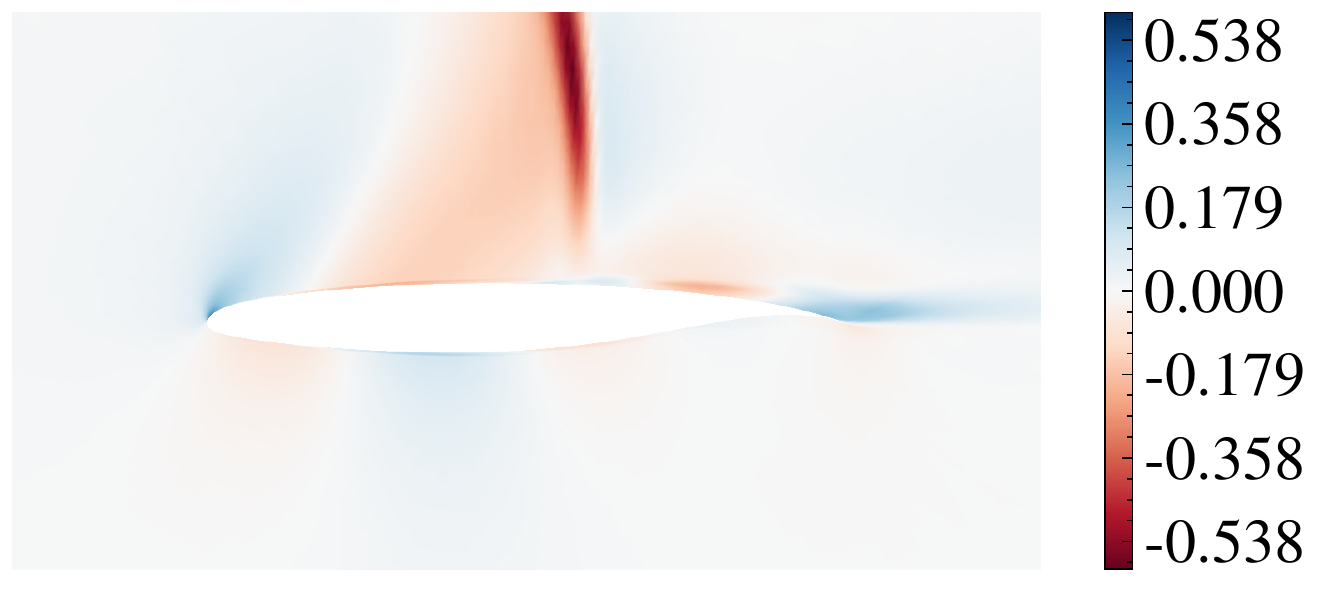}
            \put(-150,60){\textbf{z7}}
        }
	\subfigure[$\alpha=100.0$]{
		\centering
		\includegraphics[width=0.33\linewidth]{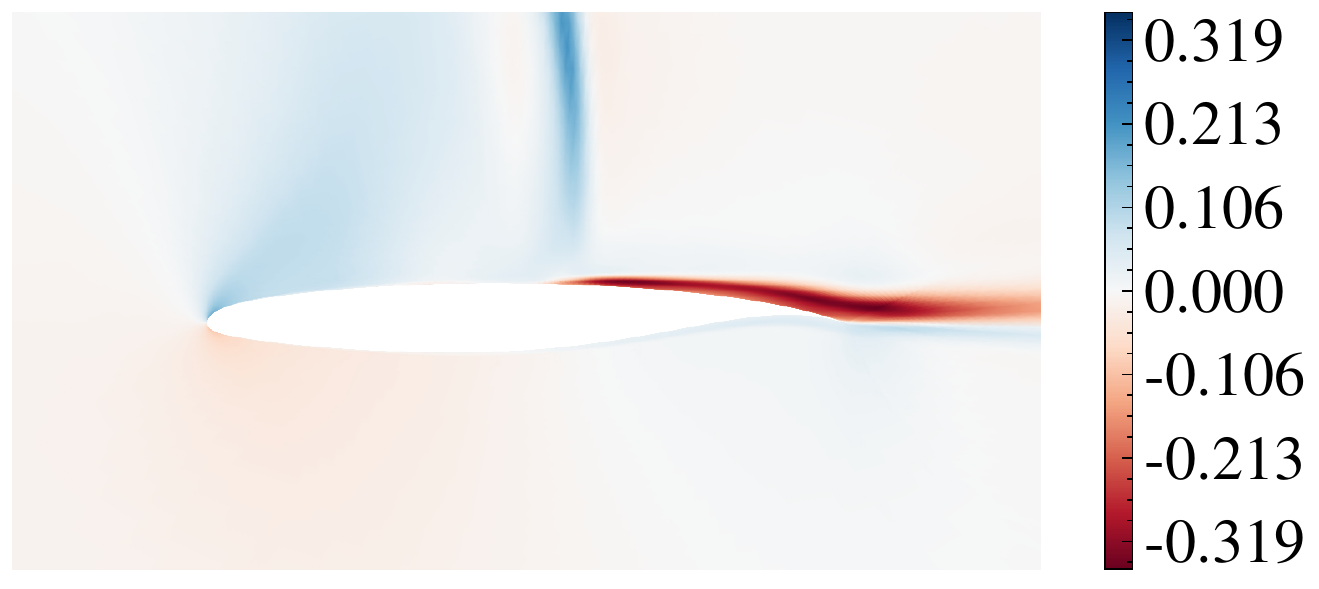}
            \put(-150,60){\textbf{z6}}
		\includegraphics[width=0.33\linewidth]{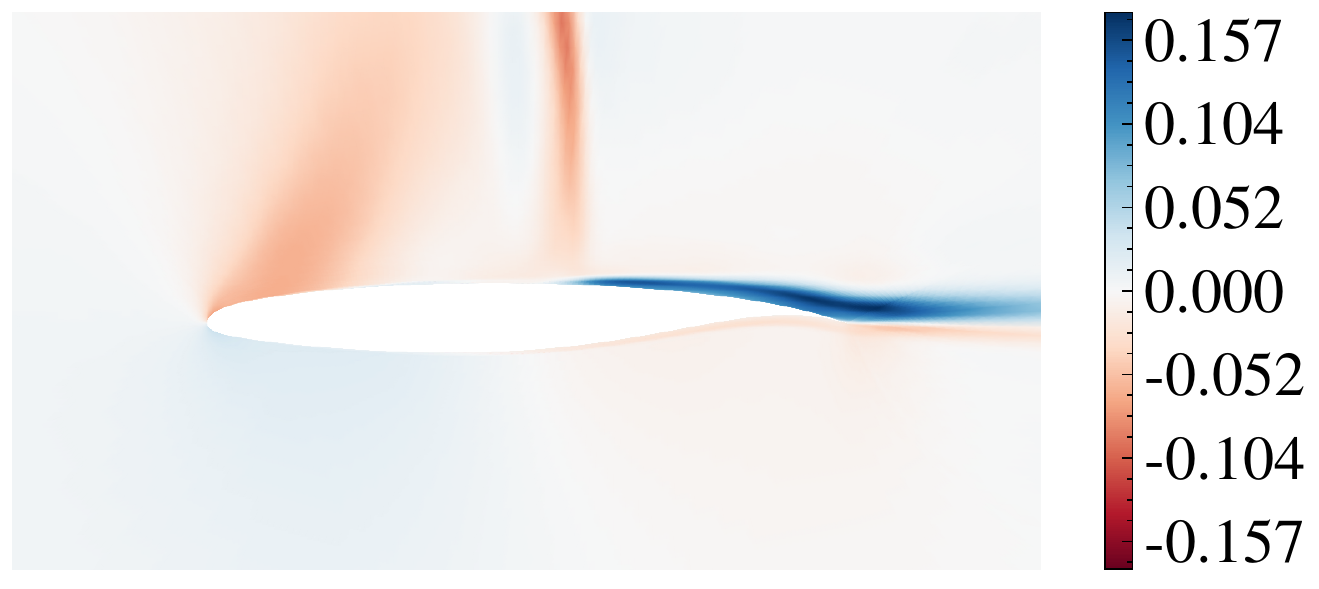}
            \put(-150,60){\textbf{z1}}
		\includegraphics[width=0.33\linewidth]{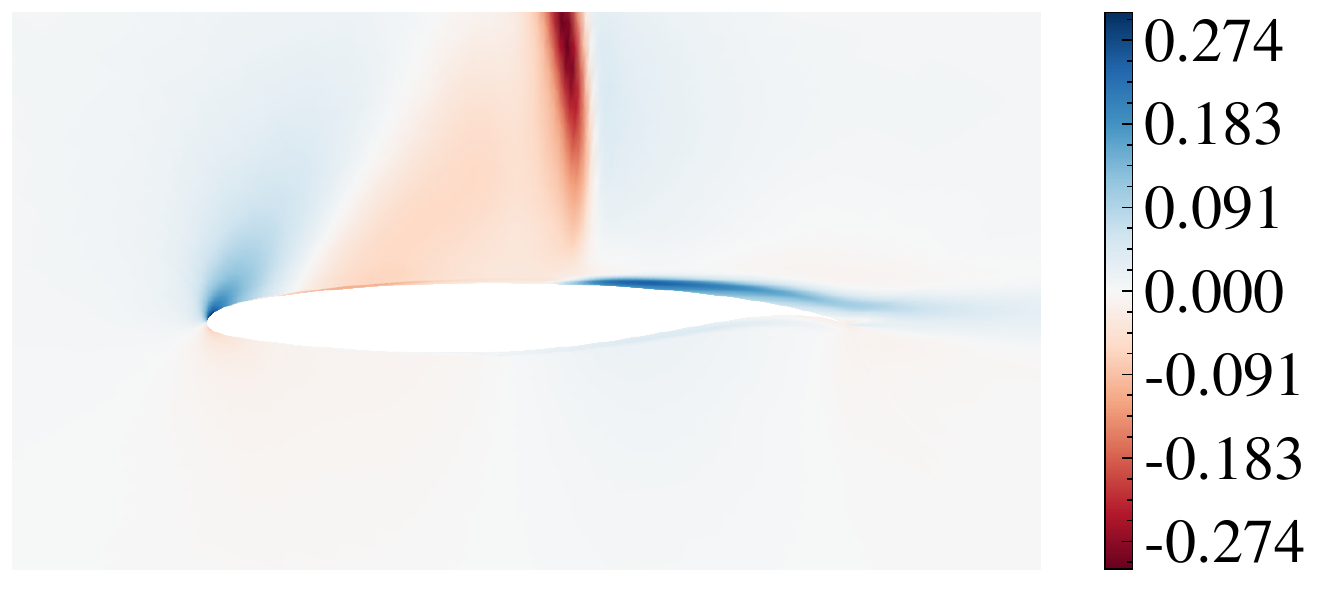}
            \put(-150,60){\textbf{z8}}
        }
	\caption{Latent traversal fields for the first three latent variables obtained with different $\alpha$}
	\label{fig:traverse1}
\end{figure}

To provide insights into how the model arrives at its decisions, identify the important factors or patterns it relies on, and increase transparency and trust in the network's results, techniques such as Class Activation Mapping (CAM)~\cite{selvaraju2017grad,ribeiro2016should} have been widely utilized. 
CAM achieves this by backpropagating the gradients of the output class to the input, highlighting the influential regions in the input image that contribute to the prediction.
However, it is important to acknowledge that CAM may suffer from limitations due to the spatial interpolation used in its generation process. This interpolation can result in the loss of precise spatial localization, leading to blurry and less accurate boundaries in the resulting CAM regions. 
For complex flow scenarios with diverse scale features and vortices, this coarseness can pose challenges in capturing fine details and regions of true significance.
In the present study, PAVAE offers a more direct and precise approach for mapping the low-dimensional latent space back to the original high-dimensional flow fields using the decoder, rather than relying on the backpropagation. 
Each latent space is visualized via a latent traversal plot, which shows the reconstruction of a latent space while keeping the remainder fixed.
The mean traversal fields of each latent space across all samples are calculated by
The mean traversal fields of each latent variable $\mathbf{z}_j$ for all the samples are calculated by
\begin{equation}
\varPhi_j = \frac{1}{N}\sum_{i=1}^{N}\text{Decoder}\left(\mathbf{z}^i_j + dz;\mathbf{z}^i_{-j} \right)-\text{Decoder}\left(\mathbf{z}^i\right),\quad j=0,1,...,9
\end{equation}
It is found that the traversal fields with various $dz$ yield $\varPhi_j$ with different magnitudes but identical flow patterns. 
The traversal fields of $dz=0.5$ for the first three dominant latent spaces are shown in Fig.~\ref{fig:traverse1}.
It is interesting to note that the traversal fields exhibit variations that correspond to distinct flow characteristics, thereby exhibiting a similarity to the modes encountered in techniques such as POD or Dynamic Mode Decomposition (DMD). Each traversal field captures a particular mode within the flow field, thereby facilitating an understanding of the impact of diverse flow characteristics on the predictions of the network. This correspondence establishes a valuable connection between the PAVAE model and established flow analysis techniques, fostering insights and enabling further exploration of the relationships between latent space representations and flow analysis methodologies.

For better comparison, Fig.~\ref{fig:traverse2} presents the latent traversal plots along the 20th grid of the normal direction obtained using $zi=\mu-3\sigma, \mu-1.5\sigma, \mu, \mu+1.5\sigma, \mu+3\sigma$.
Since the variations in the lower surface are insignificant, only the upper surface is displayed in the figure.
When $\alpha=0.0$ and $\alpha=1.0$, all three latent variables are closely related with the buffet. 
It is seen that each latent variable leads to significant velocity perturbations in the regions near the shock wave and boundary layer, which is consistent with the human expertise that transonic buffet is caused by the interaction between the separated boundary layer and the shock wave. 
However, these perturbations are coupled with each other, making it challenging to extract a specific characteristic associated with transonic buffet.
On the other hand, the variations obtained from $\alpha=10.0$ and $\alpha=100.0$ are relatively disentangled from each other.
The first and most concerned latent space is mainly concentrated in the boundary layer downstream of the shock, and the separated boundary-layer are also involved.
This finding is consistent with the mode shape of OAT15A obtained
by~\citet{crouch2019global}, and the zones that are absolutely necessary for the instability by~\citet{paladini2019various}.  
Finally, it should be noted that the dominant region identified through this analysis should not be viewed as the origin of the transonic buffet, but rather as a factor closely related to the transonic buffet that can contribute to its identification.

\begin{figure}[htpb]
	\centering
	\subfigure[$\alpha=0.0$]{
		\centering
		\includegraphics[width=\linewidth]{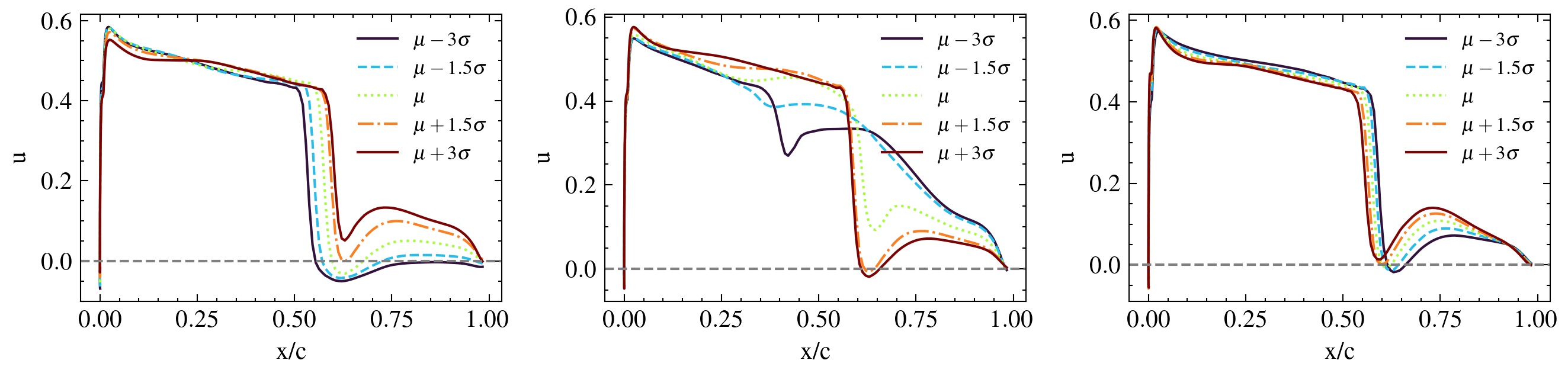}}
	\subfigure[$\alpha=1.0$]{
		\centering
		\includegraphics[width=\linewidth]{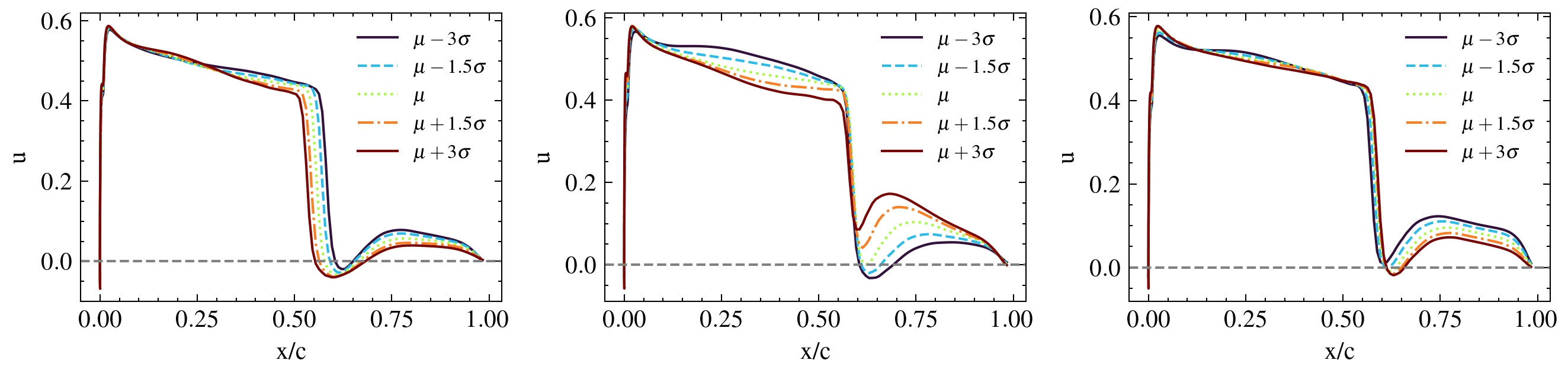}}
	\subfigure[$\alpha=10.0$]{
		\centering
		\includegraphics[width=\linewidth]{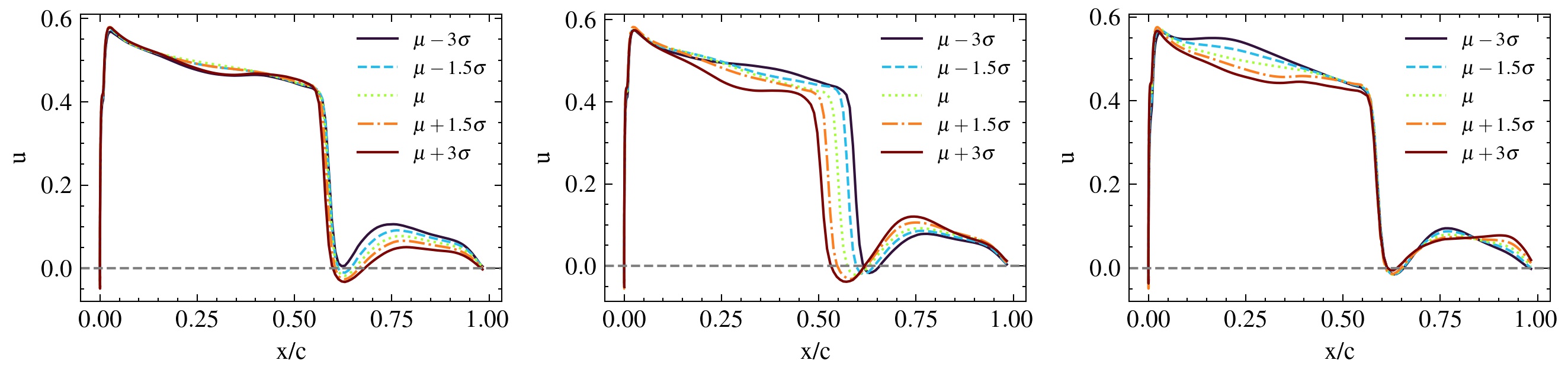}}
	\subfigure[$\alpha=100.0$]{
		\centering
		\includegraphics[width=\linewidth]{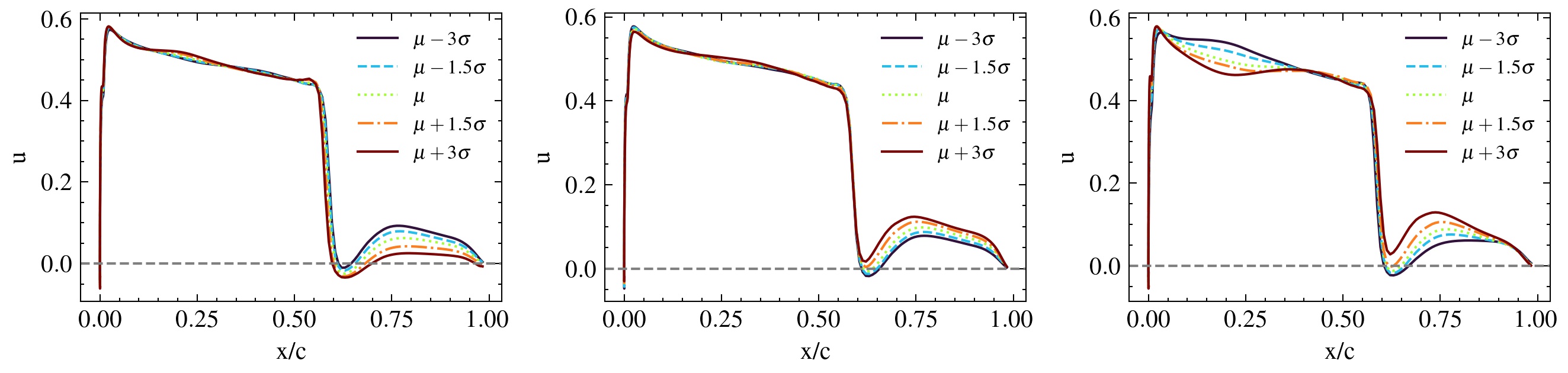}}
	\caption{Latent traversal field of upper surface along the 20th grid of the normal direction for the first three latent variables obtained with different $\alpha$}
	\label{fig:traverse2}
\end{figure}

\subsection{Buffet analysis with the active characteristics}
In the previous study, the region located in the shear layer downstream of the shock has been identified to   
strongly correlate with transonic buffet and precisely indicate the buffet state.
Accordingly, we would like to propose a physical metric based on this dominant region to perform buffet analysis. The metric allows for the formulation of a constraint function, which can predict whether a design falls within the buffet boundary, enabling the quantification of buffet onset and the application of appropriate constraints in aerodynamic shape optimization. 
For the metric to be useful in gradient-based optimization algorithms, it must be continuous and vary smoothly. 
Although the actual physical behavior is highly nonlinear, buffet onset is a gradual process, suggesting that the development of such a function is feasible.

\citet{kenway2017buffet} proposed a separation metric for transonic buffet analysis based on flow separation area and found that a 4\% cutoff of the metric achieved the best agreement with the lift curve break method. 
However, some discrepancies have been observed between their approach and the $\Delta\alpha=0.1^\circ$ method~\cite{li2022physics}. 
In this study, a statistical analysis is performed on a dataset of 26912 samples, revealing that separation area is insufficient to fully classify buffet state, with a classification accuracy rate of 92.8\%. 
The detailed distribution statistics of separation area for pre-buffet and post-buffet samples are presented in Fig.~\ref{fig:distsep}. 
The results indicate that the buffet state is unpredictable when the separation area falls within the range of $[0.002, 0.005]$.

\begin{figure}[htpb]
		\centering
		\includegraphics[width=0.45\linewidth]{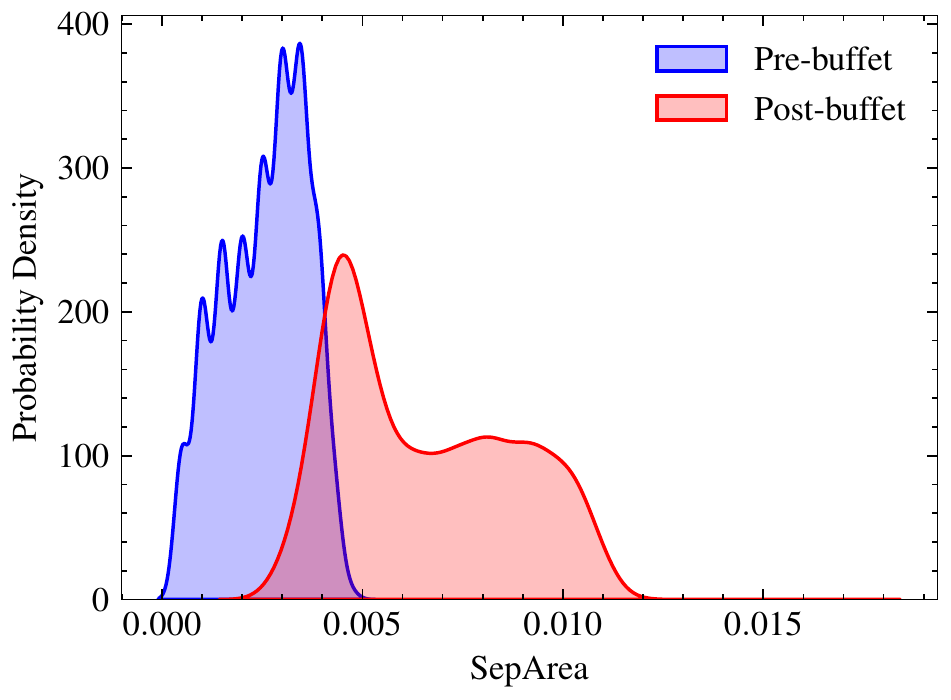}	
		\caption{Distribution statistics of separation area}
		\label{fig:distsep}	
\end{figure}

Given the significant changes in velocity observed in the shear layer downstream of the shock, particularly downstream of the attachment point, a physical metric associated with the boundary layer is proposed in this study. 
The u-velocity contours of one airfoil under pre-buffet onset and post-buffet offset states are presented in Fig.\ref{fig:compu}. 
In addition, the velocity profiles of two randomly selected airfoils at $x/c=0.8$ under different AoAs are shown in Fig.~\ref{fig:velocityprofile}.
The results demonstrate that the boundary layer downstream of the shock exhibits distinct forms with regular changes in the velocity profile as AoA increases, which can be characterized by the boundary layer thickness. 
Therefore, a displacement thickness sensor based on the velocity profile is proposed as a predictor for buffet onset. 
Displacement thickness represents the distance that the body should displace to consider the boundary layer effects in the equivalent inviscid flow, and is an appropriate indicator for the boundary layer characteristic.

\begin{figure}[htpb]
	\centering
	\subfigure[Pre-buffet]{
		\includegraphics[width=0.45\textwidth]{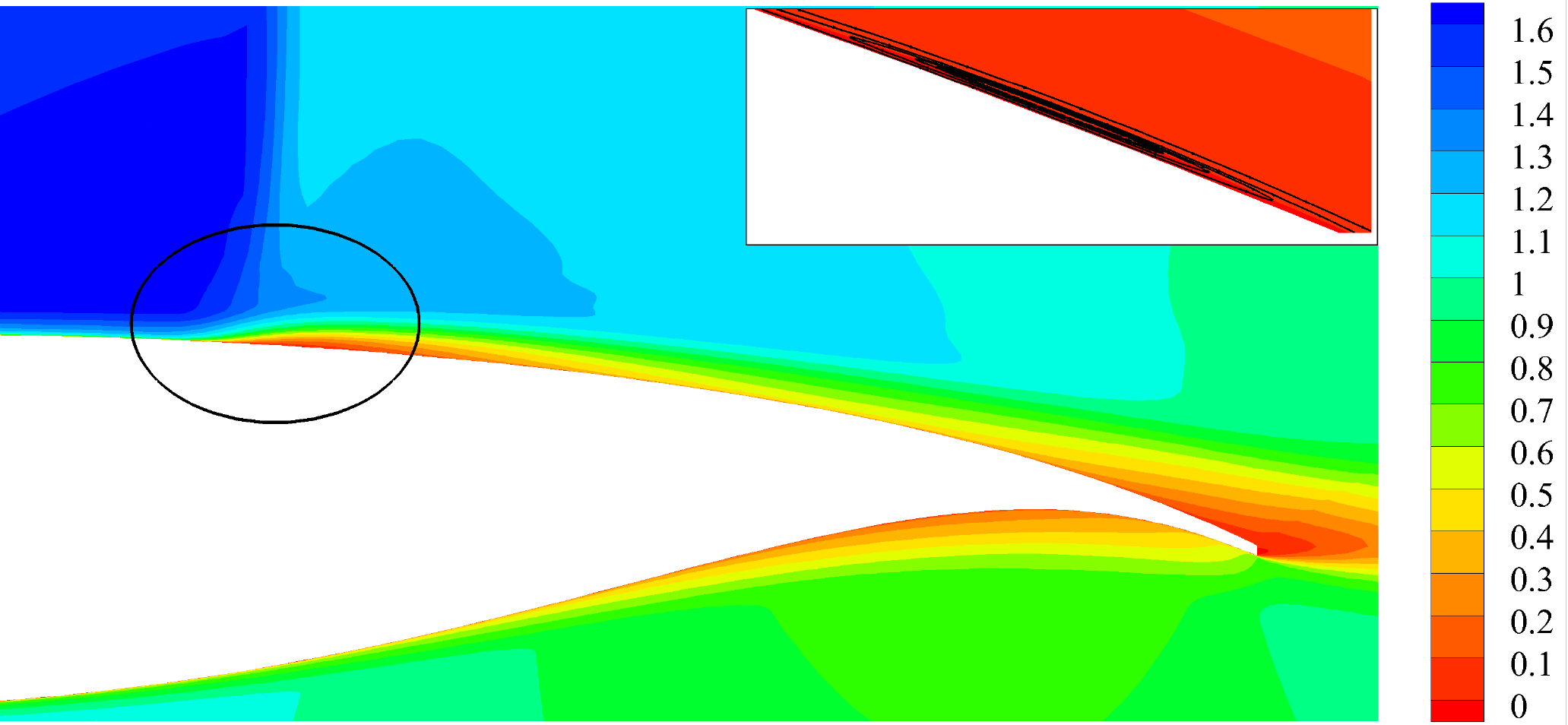}}
	\hspace{12pt}
	\subfigure[Post-buffet]{
		\includegraphics[width=0.45\textwidth]{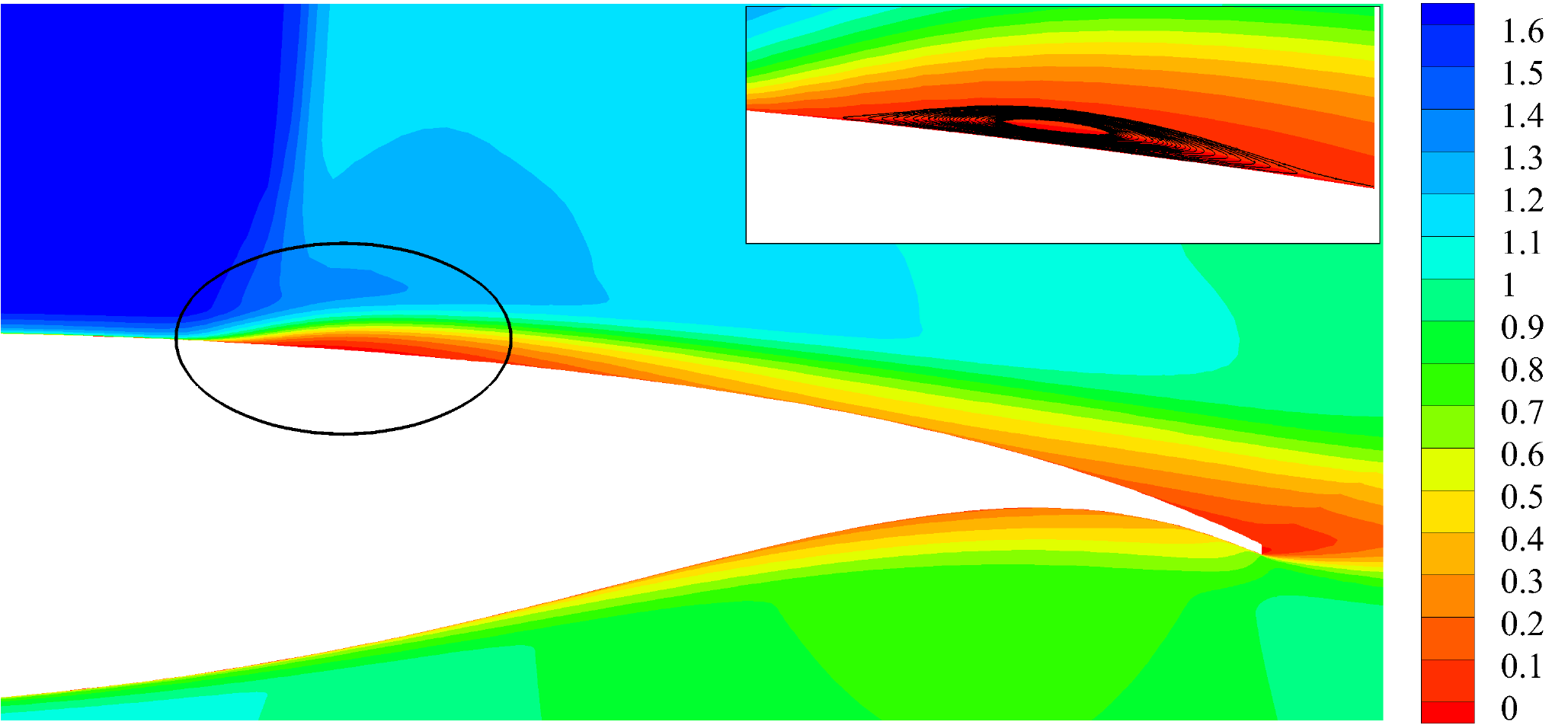}}
	\caption{U-velocity contour for one airfoil under pre-buffet onset and post-buffet offset conditions}
	\label{fig:compu}
\end{figure}

\begin{figure}[htpb]
	\centering
	\includegraphics[width=0.95\linewidth]{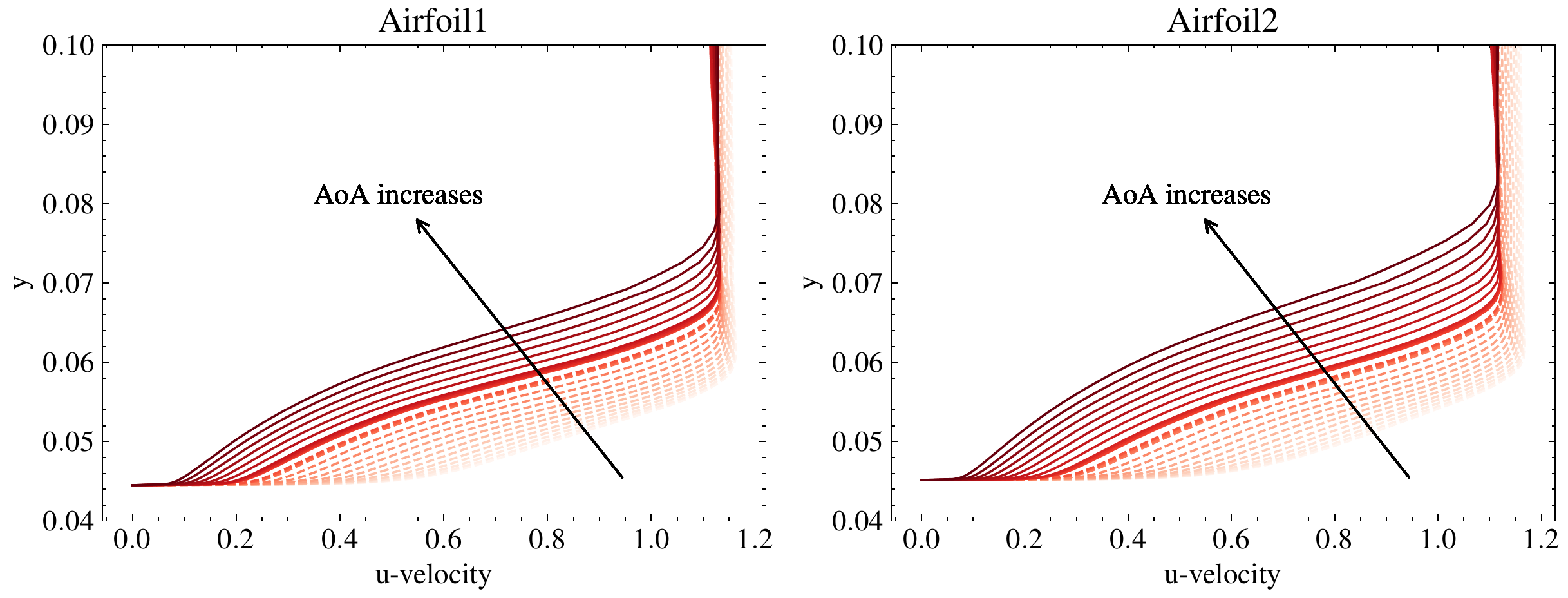}
	\caption{Velocity profiles of two random selected airfoils at different AoAs, where y indicates the  distance from the surface of the airfoil (Dash lines: Pre-buffet, Solid lines: Post-buffet)}
	\label{fig:velocityprofile}
\end{figure}

This work compares different chordwise locations of the boundary layer to determine the best criterion for classifying pre- and post-buffet states. 
Specifically, the chordwise locations of $x/c=0.75, 0.78, 0.8, 0.83$ are investigated, and the classification accuracies are reported as 96.9\%, 97.7\%, 98.5\%, and 95.8\%, respectively. 
The results reveal that the location of $x/c=0.8$ yields the best classification performance, outperforming the separation area metric. 
This finding can be supported by the distribution statistics of the maximum u-velocity in the shear layer downstream of the shock, which is presented in Fig.~\ref{fig:distxpos}. The maximum u-velocity of pre-buffet samples mainly locates at $[0.55, 0.8]$, while that of post-buffet samples mainly locates at $[0.73, 0.87]$. 
Accordingly, pre-buffet samples may exhibit larger thickness at $x/c=0.75, 0.78$ than post-buffet samples, resulting in lower accuracy than at $x/c=0.8$. 
Conversely, $x/c=0.83$ attains the lowest accuracy since the corresponding boundary layer of most samples is located downstream of the maximum u-velocity, and the corresponding discrepancies between pre- and post-samples are not as evident.

\begin{figure}[htpb]
	\centering
	\begin{minipage}[t]{0.45\textwidth}
		\centering
		\includegraphics[width=1\linewidth]{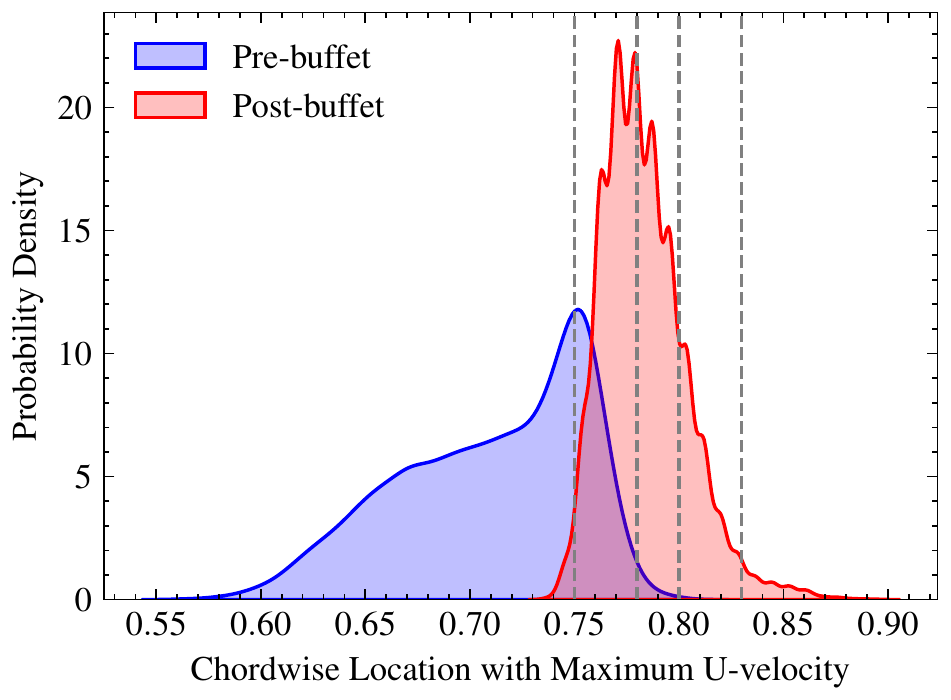}	
		\caption{Distribution statistics of chordwise location with local maximum u-velocity}
		\label{fig:distxpos}	
	\end{minipage}
	\hspace{10pt}
	\begin{minipage}[t]{0.45\textwidth}
			\centering
			\includegraphics[width=1\linewidth]{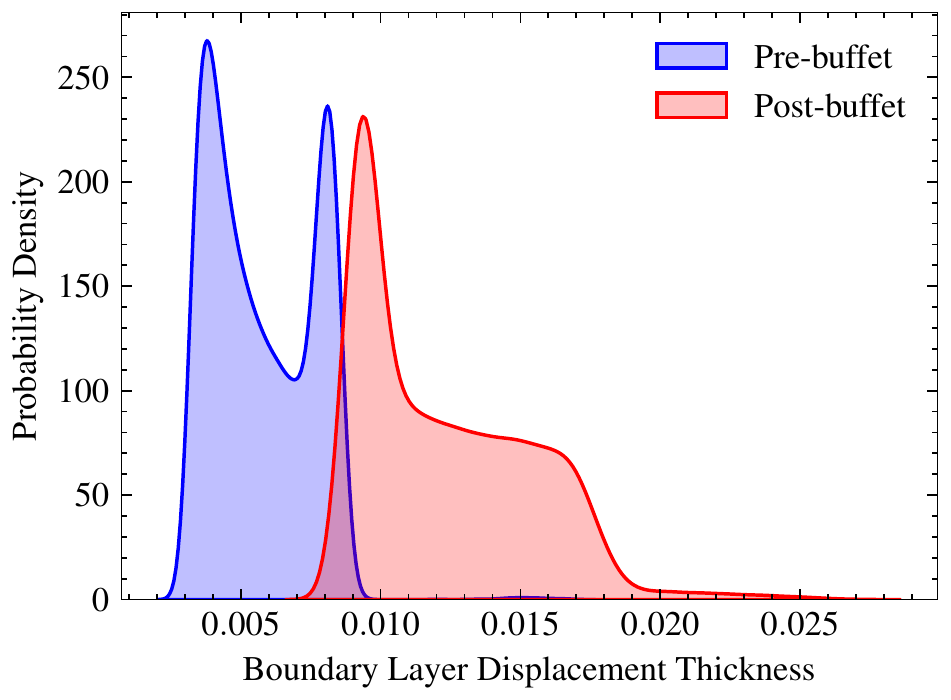}	
			\caption{Distribution statistics of displacement thickness at $x/c=0.8$}
			\label{fig:distthick}	
		\end{minipage}	
\end{figure}

\section{Conclusions}
To identify the dominant features related to transonic buffet, this paper presents a novel approach, which combines the advantages of feature extraction, flow reconstruction, and buffet prediction within a single framework.
The main new contributions of this paper can be summarized as follows:

(1) A Physics-Assisted Variational Autoencoder network is proposed. The inclusion of the buffet state as a label in the classifier enhances the capability
of the method to capture features specifically related to buffet conditions. 
By training the
network with this additional information, it becomes possible to identify and analyze the
features that are relevant to the buffet phenomenon, providing valuable insights into its underlying dynamics. Therefore, the proposed method not only performs feature extraction
and flow reconstruction, similar to conventional unsupervised learning techniques, but also
goes beyond by incorporating buffet-related features. This enriches the current understanding of flow physics by uncovering specific features and correlations that are important in predicting and analyzing transonic buffet.

(2) Four models with various weights ($\alpha=0.0, 1.0, 10.0, 100.0$) adjusting the contribution of the classifier are trained to explore the effect of buffet information on the latent space.
The high-dimensional flow field data of size $384\times192$ is compressed into ten representative features. 
From the statistical results, it is seen that the mean absolute error is below 0.0004 and the accuracy is nearly 100\% in all the four models.
Based on the comprehensive analysis, larger $\alpha$ tends to exhibit worse reconstruction performance,
whereas it is conductive to a latent space that can capture more significant differences between pre- and post-buffet samples with better classification performance.
Moreover, a rational $\alpha$ can contribute to an optimal latent space with both good reconstruction and classification performance.

(3) The contribution of each latent space obtained with various weights to the transonic buffet is analysed numerically.
10 latent variables are ranked based on their contribution and  
it is found that most of the latent variables can be regarded as inactive with nearly zero contribution to the transonic buffet.
Four models all achieve nearly 100\% accuracies when using 3D latent space with the first three latent variables involved. 
$\alpha=10.0$ and $\alpha=100.0$ realizes nearly complete differentiation using just 1D latent space.
Then, the dominant regions related with transonic buffet is obtained through the latent traversal plot of each latent variable.
The results show that the first and most concerned latent space is mainly concentrated in the boundary layer downstream of the shock.

(4) A displacement thickness sensor is proposed as a predictor for buffet onset based on the dominant region.
It is demonstrated that the classification accuracy of the buffet state based on this metric reaches 98.5\%, which is much higher than 92.8\% based on the separation area.
Once built, this metric has a good generalizable relationship applies to various aerodynamic shapes.

Overall, this study presents a promising approach that leverages big data and machine learning to analyse complex flow phenomena. 
Notably, the approach is feasible for interpreting the "black box" neural network like CAM, offering valuable insights for the AI research community and potentially leading to further advancements in the field. 
Future work will concentrate on utilizing high-fidelity models, such as large-eddy simulations, to provide the original field data and label the training data, ultimately enhancing the accuracy and reliability of the proposed approach.

\section*{Acknowledgments}
This study was supported by the Shanghai Sailing Program (No. 21YF1459400).

\bibliography{buffetcomp}

\end{document}